\documentclass{aastex63}

\usepackage{wrapfig}

\submitjournal{ApJ}

\shorttitle{Nitrogen Dioxide Pollution as a Signature of Extraterrestrial Technology}
\shortauthors{Kopparapu et al.}
\graphicspath{{./}{figures/}}
\begin{document}

\title{Nitrogen Dioxide Pollution as a Signature of Extraterrestrial Technology}

\correspondingauthor{Ravi Kopparapu}
\email{ravikumar.kopparapu@nasa.gov}

\author[0000-0002-5893-2471]{Ravi Kopparapu}
\affiliation{NASA Goddard Space Flight Center \\
8800 Greenbelt Road \\
Greenbelt, MD 20771, USA}

\author{Giada Arney}
\affiliation{NASA Goddard Space Flight Center \\
8800 Greenbelt Road \\
Greenbelt, MD 20771, USA}
%\author{August Muench}
%\affiliation{American Astronomical Society \\
%1667 K Street NW, Suite 800 \\
%Washington, DC 20006, USA}

%\collaboration{1}{(AAS Journals Data Scientists collaboration)}

\author[0000-0003-4346-2611]{Jacob Haqq-Misra}
\affiliation{Blue Marble Space Institute of Science,\\
Seattle, WA, USA}

\author[0000-0002-0746-1980]{Jacob Lustig-Yaeger}
\affiliation{Johns Hopkins University Applied Physics Laboratory,\\ 
Laurel, MD 20723, USA}

\author[0000-0002-2662-5776]{Geronimo Villanueva}
\affiliation{NASA Goddard Space Flight Center \\
8800 Greenbelt Road \\
Greenbelt, MD 20771, USA}

%\affiliation{AAS Journals Associate Editor-in-Chief}
%\nocollaboration{1}

%\author{Amy Hendrickson}
%\altaffiliation{AASTeX v6+ programmer}
%\affiliation{TeXnology Inc.}

%\collaboration{1}{(LaTeX collaboration)}

%\author{Julie Steffen}
%\affiliation{AAS Director of Publishing}
%\affiliation{American Astronomical Society \\
%1667 K Street NW, Suite 800 \\
%Washington, DC 20006, USA}

%\author{Scott Chernoff}
%\affiliation{IOP Publishing, Washington, DC 20005}

%\nocollaboration{2}

%% Note that the \and command from previous versions of AASTeX is now
%% depreciated in this version as it is no longer necessary. AASTeX 
%% automatically takes care of all commas and "and"s between authors names.

%% AASTeX 6.3 has the new \collaboration and \nocollaboration commands to
%% provide the collaboration status of a group of authors. These commands 
%% can be used either before or after the list of corresponding authors. The
%% argument for \collaboration is the collaboration identifier. Authors are
%% encouraged to surround collaboration identifiers with ()s. The 
%% \nocollaboration command takes no argument and exists to indicate that
%% the nearby authors are not part of surrounding collaborations.

%% Mark off the abstract in the ``abstract'' environment. 
\begin{abstract}
Nitrogen dioxide (NO$_{2}$) on Earth today has biogenic and anthropogenic sources. During the COVID-19 pandemic, observations of global NO$_{2}$ emissions have shown significant decrease in urban areas. Drawing upon this example of NO$_{2}$ as an industrial byproduct, we use a one-dimensional photochemical model and synthetic spectral generator to assess the detectability of NO$_{2}$ as an atmospheric technosignature on exoplanets.  We consider cases of an Earth-like planet around Sun-like, K-dwarf and M-dwarf stars. We find that NO$_{2}$ concentrations increase on planets around cooler stars due to less short-wavelength photons that can photolyze NO$_{2}$. In cloud-free results, present Earth-level NO$_{2}$ on an Earth-like planet around a Sun-like star
%K-dwarf 
at 10pc can be detected with SNR $\sim 5$ within $\sim 400$ hours with a 15 meter LUVOIR-like telescope when observed in the $0.2 - 0.7\mu m$ range where NO$_{2}$ has a strong absorption. However, clouds and aerosols can reduce the detectability and could mimic the NO$_{2}$ feature.
%Direct imaging observations by ground-based 30m class telescopes can detect $5 \times$ present Earth-level NO$_{2}$ within 10 hours of observations with SNR $> \sim 3$.
Historically, global NO$_{2}$ levels were 3x higher, indicating the capability of detecting a 40-year old Earth-level civilization.
%By comparison, a habitable planet orbiting a K-dwarf star would only need 2 times the NO$_{2}$ abundance to be detectable with the same 100 hour observation time. This is due to less availability of short-wavelength photons for a K-dwarf that can photolyze NO$_{2}$. Detection of Earth-level NO$_{2}$ concentrations around a Sun-like star at a SNR $\sim 3$ would take nearly 5000 hours of LUVOIR-A time, for a planet at 10\,pc {\bf when observed in the $0.2 - 0.7\mu m$ range where NO$_{2}$ has a strong absorption}. 
Transit and direct imaging observations to detect infrared spectral signatures of NO$_{2}$ on habitable planets around M-dwarfs would need several 100s of hours of observation time, both due to weaker NO$_{2}$ absorption in this region, and also because of masking features by dominant H$_{2}$O and CO$_{2}$ bands in the infrared part of the spectrum. Non-detection at these levels could be used to place upper limits on the prevalence of NO$_{2}$ as a technosignature.
%as future missions continue the spectroscopic characterization of exoplanet atmospheres. 
\end{abstract}

%% Keywords should appear after the \end{abstract} command. 
%% See the online documentation for the full list of available subject
%% keywords and the rules for their use.

%% From the front matter, we move on to the body of the paper.
%% Sections are demarcated by \section and \subsection, respectively.
%% Observe the use of the LaTeX \label
%% command after the \subsection to give a symbolic KEY to the
%% subsection for cross-referencing in a \ref command.
%% You can use LaTeX's \ref and \label commands to keep track of
%% cross-references to sections, equations, tables, and figures.
%% That way, if you change the order of any elements, LaTeX will
%% automatically renumber them.
%%
%% We recommend that authors also use the natbib \citep
%% and \citet commands to identify citations.  The citations are
%% tied to the reference list via symbolic KEYs. The KEY corresponds
%% to the KEY in the \bibitem in the reference list below. 

\section{Introduction} \label{sec:intro}
Over the last 25 years, more than 4000 exoplanets have been discovered\footnote{\url{https://exoplanetarchive.ipac.caltech.edu/}} from both ground and space-based surveys. We are now entering into an era of exoplanet atmospheric characterization, with the soon to be launched James Webb Space Telescope ({\it JWST}), Atmospheric Remote-sensing Infrared Exoplanet Large-survey (ARIEL) space telescope, and large ground-based observatories such as the European Extremely Large Telescope (E-ELT), the Thirty Meter Telescope (TMT), and the Giant Magellan Telescope (GMT). The first detection of an exoplanet atmosphere was on a gas giant planet, HD 209458b, in 2001 \citep{dave2002}. Since then,  atmospheres have been detected on exoplanets spanning a wide range of planetary parameter space, and observers are continuing to push the limits towards smaller worlds  \citep{paper12019,paper22019}. The ongoing discovery of exoplanet atmospheres has raised the prospect of eventually identifying potentially habitable planets, as well as the possibility of finding one that may also be inhabited. As a result, the characterization and detection of ``biosignatures,''---remote observations of atmospheric spectral features that could potentially indicate signs of life on an exoplanet---has received recent attention as an area of priority for astrobiology\footnote{\url{https://www.nationalacademies.org/our-work/astrobiology-science-strategy-for-the-search-for-life-in-the-universe}} \citep{seager2012astrophysical, kaltenegger2017characterize, Schwieterman2018, Meadows2018, Catling2018, Walker2018, Fujii2018, o2019expanding, lammer2019role, grenfell2017review}.

Similar to biosignatures, ``technosignatures'' refer to any observational manifestations of extraterrestrial technology that could be detected or inferred through astronomical searches. As discussed in the 2018 NASA technosignatures workshop report \citep{tech2018}:  ``Searches for technosignatures are logically continuous with the search for biosignatures as part of astrobiology. As with biosignatures, one must proceed by hypothesizing a class of detectable technosignatures, motivated by life on Earth, and then designing a search for that technosignature considering both its detectability and its uniqueness.''  Although the science of atmospheric technosignatures is less developed compared to atmospheric biosignatures, a wide class of possible technosignatures have been suggested in the literature that include waste heat \citep{dyson60,GHAT2,kuhn2015global,carrigan09b}, artificial illumination \citep{Schneider2010,Loeb11,Kipping16}, artificial atmospheric constituents \citep{Schneider2010,Lin2014,Stevens16}, artificial surface constituents \citep{Lingam17}, stellar ``pollution'' \citep{shklovskii1966,Whitmire80,Stevens16}, non-terrestrial artifacts \citep{BRACEWELL1960,freitas1980search,Rose2004,JR2012}, and megastructures \citep{dyson60,Arnold05,Forgan13,GHAT4}. This breadth of topics reflects the scope of possibilities for detecting plausible technosignatures, although the sophistication of technosignature science remains in its infancy compared to the rapidly evolving field of biosignatures \citep{wright2019,jacob2020}. 

The history of life on Earth provides a starting point in the search for biosignatures on exoplanets  \citep{Josh2018, palle2018}, with the various stages of Earth's evolution through the Hadean (4.6 -4 Gyr), Archean (4 - 2.5 Gyr),  Proterozoic (2.5 - 0.54 Gyr), and Phanerozoic (0.54 Gyr - present) eons representing atmospheric compositions to use as examples of spectral signatures of an inhabited planet. The use of Earth's history as an example does not imply that these biosignatures will necessarily be the most prevalent in the galaxy, but instead this approach simply represents a place to begin based on the one known example of life. By extension, the search for technosignatures likewise can consider Earth's evolution into the Anthropocene epoch \citep{crutzen2006anthropocene,lewis2015defining,frank2017earth} as a template for future observing campaigns that seek to detect evidence of extraterrestrial technology. For instance, \cite{Lin2014} discussed the possibility of detecting tetrafluoromethane (CF$_{4}$) and trichlorofluoromethane (CCl$_{3}$F) signatures in the atmospheres of transiting Earth-like planets around white dwarfs with JWST, which could be detectable if these compounds are present at 10 times the present Earth level. 
These chlorofluorocarbons (CFCs) are produced by industrial processes on Earth, so their detection in an exoplanet atmosphere could be strong evidence for the presence of extraterrestrial technology. This approach does not insist that CFCs or other industrial gases found on Earth will necessarily be the most prevalent technosignature in the galaxy, but it represents a place to begin defining observables and plausible concepts for technosignatures based upon the one known example of technological civilization.

In this study, we explore the possibility of NO$_{2}$ as an atmospheric technosignature. Some NO$_{2}$ on Earth is produced as a byproduct of combustion, which suggests the possibility of scenarios in which larger-scale production of NO$_{2}$ is sustained by more advanced technology on another planet. Detecting high levels of NO$_{2}$ at levels above that of non-technological emissions found on Earth could be a sign that the planet may host active industrial processes. In section \ref{sec:production}, we describe the production reactions of NO$_{2}$ and use a 1-dimensional photochemical model to obtain self-consistent mixing ratio profiles of nitrogen oxide compounds, on a planet orbiting a Sun-like star, a K6V spectral type (T$_{eff} = 4600$K), and the two  M-dwarf stars AD Leo (T$_{eff} = 3390$K) and Proxima Centauri (T$_{eff} = 3000$K). Using these photochemical results, in section \ref{sec:detectability} we calculate the observability of strongest NO$_{2}$ features between $0.2 - 0.7\mu m$ and between $1 - 10 \mu$m using a spectral generation model to produce geometric albedo and transit spectra of planets with various facilities like LUVOIR-15m, JWST and OST. In section \ref{sec:discussion}, we discuss the implications of these observations, concluding in section \ref{sec:conclusion}.

\section{Production of Nitrogen Dioxide}\label{sec:production}

Nitrogen oxides (NO$_{x}$ = NO + NO$_{2}$) are among the main pollutants in industrialized locations on the globe.  The non-anthropogenic pathways for the production of NO$_{x}$ can be either due to emission from soils and wildfires, or produced in the troposphere by lightning.\footnote{The troposphere is the lowermost atmospheric layer from the surface up to 10-18 km, highest at the tropics and lowest near the poles during winter. The pressure and temperature decrease with altitude, with global averages of 289\,K and 1.013 millibar (mb) at the surface, and around 210K and 150\,mb at a height of 15\,km.} The primary biogenic source of NO$_{x}$ is bacteria in soil through nitrification (i.e bacteria converting ammonia to nitrite and nitrate compounds), or dentrification (process of reducing nitrate and nitrite to gaseous forms of nitrogen such as N$_{2}$ or N$_{2}$O).
The estimated worldwide biogenic and lightning emissions of NO$_{x}$ compounds are $\sim 10.6$ Tera gram per year as N (Tg(N) yr$^{-1}$, (Table 1,  \cite{Holmes2013}). Lightning contributes about 5 Tg(N) yr$^{-1}$, which translates to $6 \times 10^{8}$ NO molecules/cm$^{2}$/s \citep{Harman2018}. 

On the other hand, NO$_{x}$ compounds are also emitted from anthropogenic sources of combustion processes such as vehicle emissions and fossil-fueled power plants. The role of this industrial production was noted during the COVID-19 pandemic, when global concentrations of NO$_{2}$ were observed to decrease between $20 -40\%$  over urban areas \citep{bauwensimpact}. Indeed, these emissions dominate the production of NO$_{x}$ compounds in the troposphere more than the biogenic sources with an estimated rate of 32 Tg(N) yr$^{-1}$ \citep{Holmes2013}.  NO$_{2}$ poses harmful health effects that could cause impairment of lung function and respiratory problems \citep{Faustini2014}.  Typical concentrations of NO$_{2}$ range from 0.01 ppb (parts per billion) to $\sim 5$ ppb depending upon the urbanization with the higher number correlating to urban areas \citep{lamsal2013scaling}. The presence of NO$_{x}$ in the lower troposphere leads to a complex chemistry that results in the formation of ozone (O$_{3}$), which is a harmful pollutant in the troposphere and a greenhouse gas. NO$_{x}$ mixing ratios in excess of $10^{-7}$ would cause severe damage to the O$_{3}$ layer and could result in either a climatic warming or cooling, depending upon the amount of NO$_{2}$ present \citep{KA1985}.

The sinks and sources for NO$_{2}$ in the troposphere ($\le$ 20km) are governed by the following reactions.
NO$_{2}$ photolysis is dominant in the wavelength range of $290 - 420$nm (See \cite{kraus1998}, Fig. 2).  The lower limit is set by the available solar UV-intensity and the upper wavelength limit is determined by the fall-off in the photodissociation cross-section.
This NO$_{2}$  photolysis produces ground state atomic oxygen, O($^{3}$P), along with NO:
    \begin{eqnarray}
    \mathrm{NO_{2} + h \nu (< 420nm)} &\rightarrow& \mathrm{O(^{3}P) + NO}.
    \label{eq:NO2reaction}
    \end{eqnarray}
The O($^{3}$P) then can combine with an oxygen molecule to form ozone,
    \begin{eqnarray}
    \mathrm{O(^{3}P) + O_{2} + M} &\rightarrow& \mathrm{O_{3} + M},
    \end {eqnarray}
which gets destroyed by reoxidizing nitric oxide to nitrogen dioxide:
    \begin{eqnarray}
    \mathrm{NO} + \mathrm{O_{3}} &\rightarrow& \mathrm{NO_{2}} + \mathrm{O_{2}}.
    \label{eq:NO2ozone}
    \end{eqnarray}
NO also reacts with atomic oxygen (O) and the hydroperoxy radical (HO$_{2}$) to generate NO$_{2}$,
\begin{eqnarray}
\mathrm{NO} + \mathrm{O} + \mathrm{M} &\rightarrow& \mathrm{NO_{2} + \mathrm{M}} \\
\mathrm{NO} + \mathrm{HO_{2}} &\rightarrow& \mathrm{NO_{2}} + \mathrm{OH}.
\end{eqnarray}
However, these production mechanisms of NO$_{2}$ are counteracted when NO$_{2}$ reacts again with atomic oxygen to recreate NO:
\begin{eqnarray}
\mathrm{NO_{2}} + \mathrm{O} &\rightarrow& \mathrm{NO} + \mathrm{O_{2}},
\label{eq:no2_o}
\end{eqnarray}
The above reactions just cycle between NO and NO$_{2}$, so NOx is conserved. However,  NO$_{2}$ also reacts with the OH radical to form nitric acid (HNO$_{3}$), which eventually is removed from the atmosphere by rainout, which is a loss process for NOx:
\begin{eqnarray}
\mathrm{NO_{2} + OH + M} &\rightarrow& \mathrm{HNO_{3} + M}.
\end{eqnarray}

Reactions \ref{eq:NO2ozone} and \ref{eq:no2_o} form a catalytic cycle to destroy ozone with the net reaction,
\begin{eqnarray}
\mathrm{O_{3}} + \mathrm{O} \rightarrow 2 \mathrm{O_{2}}
\end{eqnarray}
indicating that high stratospheric NOx can lead to ozone depletion.

To study the steady-state abundances of NO$_{x}$ compounds in Earth-like atmospheres, we used a 1-D photochemical model \citep[described in][]{Arney2016, Arney2019}, which is part of a coupled climate-photochemistry model called `Atmos.'\footnote{https://github.com/VirtualPlanetaryLaboratory/atmos} The photochemical model is originally based on the one described in \citet{Kasting79} and has been updated extensively over the years and applied to various planetary and exoplanetary conditions \citep[e.g.][]{Segura2005, Kopparapu2012,  Domagal-Goldman2014, Harman2015, Harman2018, Lincowski2018}. The model version used here has been updated to correct the deficiencies identified in \citet{Ranjan2020}, and the public version of the model is planned to be updated.   This model solves a set of nonlinear, coupled ordinary differential equations for the mixing ratios of all species at all heights using the reverse Euler method. The method is first order in time and uses second-order centered finite differences in space. The vertical grid has 200 altitude levels, ranging from 0 km (lower boundary) to 100 km (upper boundary). The version used here includes updates described in \citet{Lincowski2018} and includes 72 chemical species involved in 309 reactions to represent a modern Earth-like planet. We considered a Sun-like star, a K6V stellar spectral type, and two M-stars (AD Leo and Proxima Centauri) in this study. For the Sun-like star we used the \citet{CK2010} model; for the K6V star, we used the spectrum of HD 85512 from the Measurements of the Ultraviolet Spectral Characteristics of Low-mass Exoplanetary Systems (MUSCLES) treasury survey \citep{France2016, Loyd2016, Youngblood2016}; for AD Leo and Proxima Centauri, we used stellar spectra described in \citet{Segura2005} and \citet{Meadows2018b}, respectively. Planets around the other stars are placed at the Earth-equivalent flux distance. 
        
For each Earth-like planet around its host star, we ran the model to steady state to obtain the mixing ratio profiles of all gaseous species, including NO$_{2}$. 
We have used a surface NO$_2$ molecular flux of $8.64 \times 10^{9}$ molecules/cm$^{2}/s$ as the standard Earth-level (1x) flux in our simulations. This number comes from converting the estimated rate of $32$ Tg(N) yr$^{-1}$ anthropogenic NO$_{x}$ compound emissions  in the troposphere\footnote{$32 \times 10^{12}$ (gram/year) /($1.67\times 10^{-24}$ gram $\times 14 \times 4\pi (6.32 \times 10^{6} m)^{2} \times 365 \times 24 \times 3600$s)  $\sim 8.64 \times 10^{9}$ molecules/cm$^{2}/s$ } to the molecular flux. We also include a fixed biogenic flux of NO as $1.0 \times 10^{9}$ molecules/cm$^{2}/s$, kept constant across all simulations. Because we do not increase the flux of NO alongside NO$_2$, our simulations may be regarded as somewhat conservative. Other fixed boundary conditions of N-bearing species include: a flux of $1.53 \times 10^{9}$ for N$_2$O, a mixing ratio of 0.78 for N$_2$, and fixed deposition velocities of $2.1\times 10^{-1}$ for HO$_2$NO$_2$ and HNO$_3$. 
%We then scaled up the standard NO$_{2}$ mixing ratios for each star by a factor of 2x, 5x, 10x, 30x to explore atmospheres with greater  NO$_{2}$ abundance compared to modern Earth. While this is not exactly self-consistent with emitted fluxes of the gas (see Discussion section), we leave detailed studies of these effects to future analyses.

\begin{figure}
    \gridline{\fig{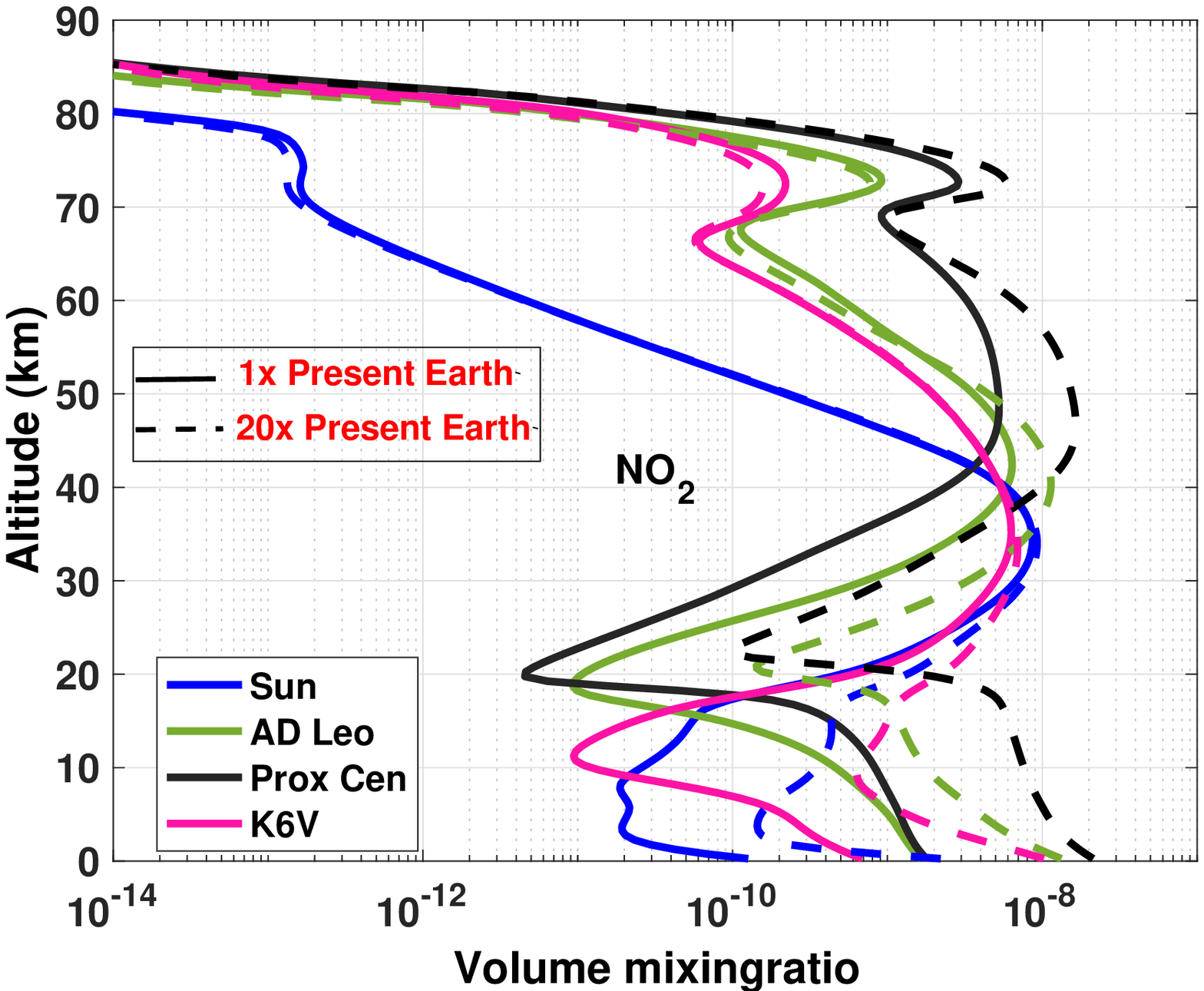}{0.49\textwidth}{(a)}
          \fig{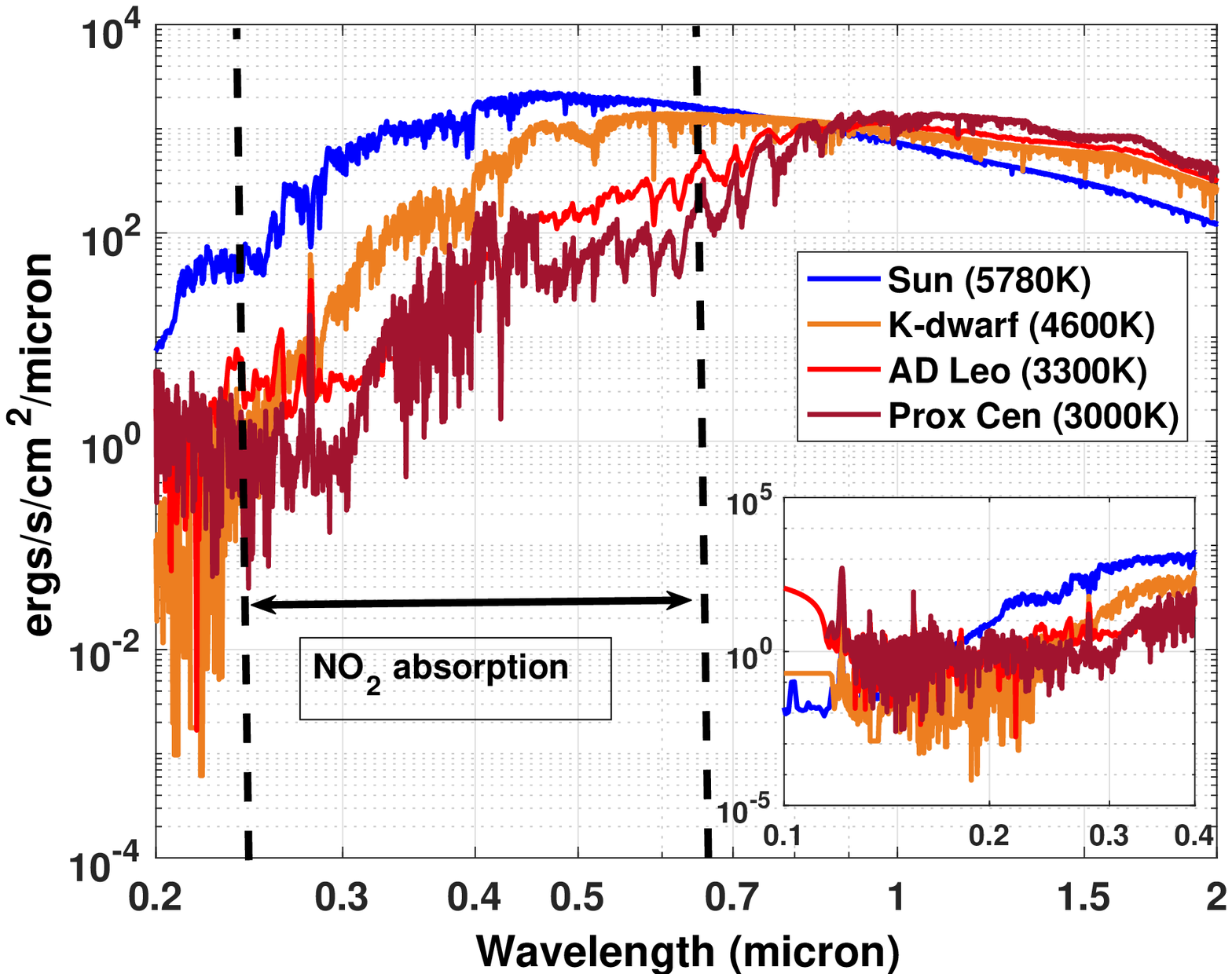}{0.49\textwidth}{(b)}
          }
    \caption{ (a): Mixing ratio profiles of  NO$_{2}$ around stars of different spectral types   on an Earth-like planet with 1x  (solid) and  20x (dashed) present Earth NO$_{2}$ fluxes. Below the troposphere ($\sim 20$km), NO$_{2}$ concentration is higher on planets around cooler stars compared to the Sun because the destruction of NO$_{2}$ is an order of magnitude more efficient around a Sun-like star due to the availability of photons of wavelength between $0.29 - 0.42 \mu$m that penetrate to the troposphere. See inset in panel b.  (b) Spectral energy distribution of stellar spectral types used in this study, indicating wavelength region of strongest NO$_{2}$ absorption. The inset shows the UV/Visible region where NO$_{2}$ photolysis happens.     } 
    \label{fig:vmr}
\end{figure}

Results from our 1-D photochemical model are shown in Fig. \ref{fig:vmr}, panel (a). This plot shows the NO$_{2}$ volume mixing ratio profiles of an Earth-like planet around four stellar spectral types we considered in this study: the Sun (blue), AD Leo (green), Proxima Centauri (black) and the K6V star (magenta). Two end member concentrations are shown: The standard Earth level flux of $8.64 \times 10^{9}$ molecules/cm$^{-2}$/s (1x, solid curves), and a flux of $172 \times 10^{9}$ molecules/cm$^{-2}$/s (20x, dashed curves).
%only those species that dominate the NO$_{2}$ production and loss rates. 
The corresponding stellar spectra are shown in the right panel (b), highlighting the wavelength region of  strongest NO$_{2}$ absorption.  As shown in this figure, the hotter stars provide more photons between 0.25 to 0.65 $\mu$m, which increases the rate of NO$_{2}$ photolysis (Eq. \ref{eq:NO2reaction}, and also Table \ref{table1}). 

\begin{figure}
    \gridline{\fig{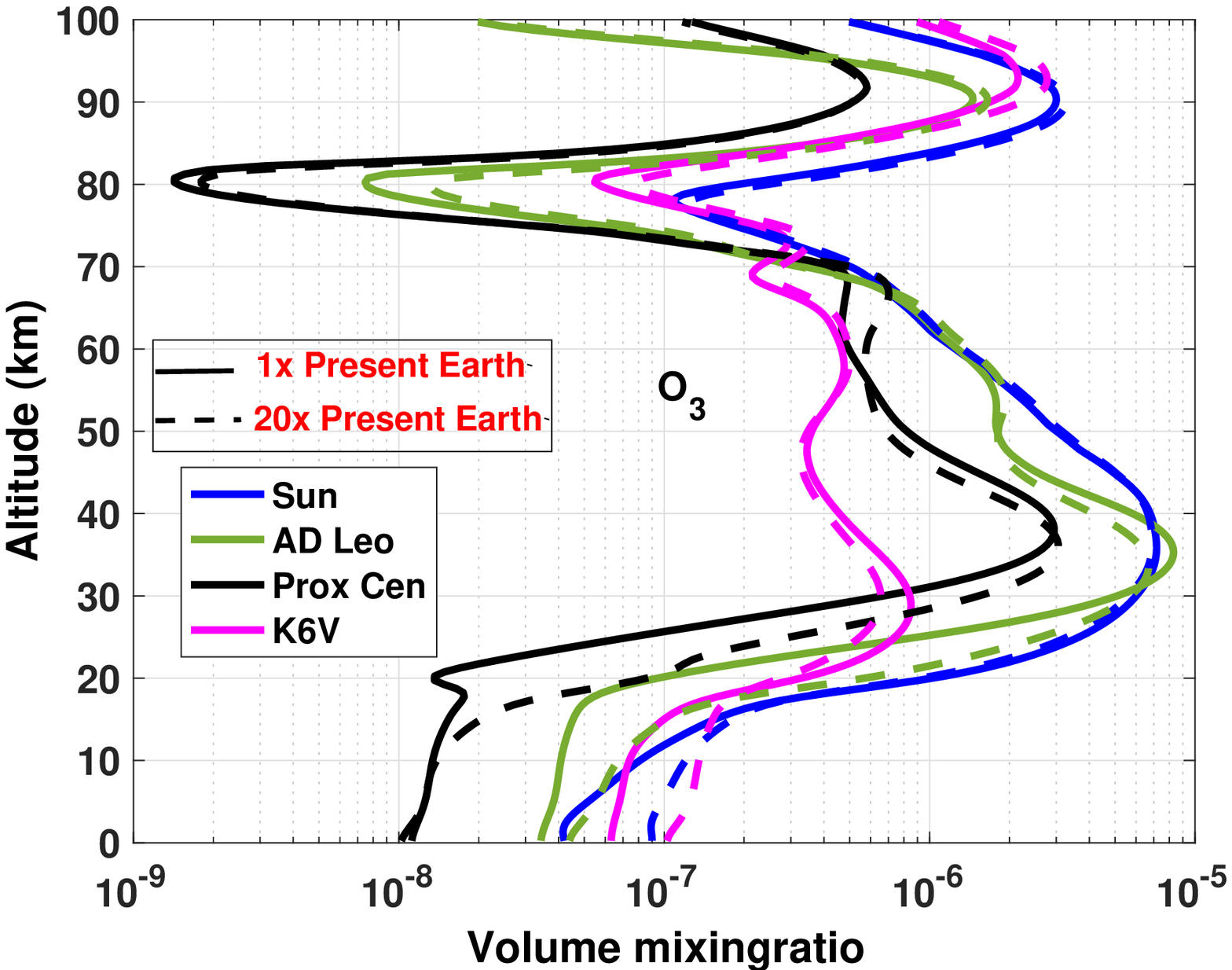}{0.49\textwidth}{(a)}
          \fig{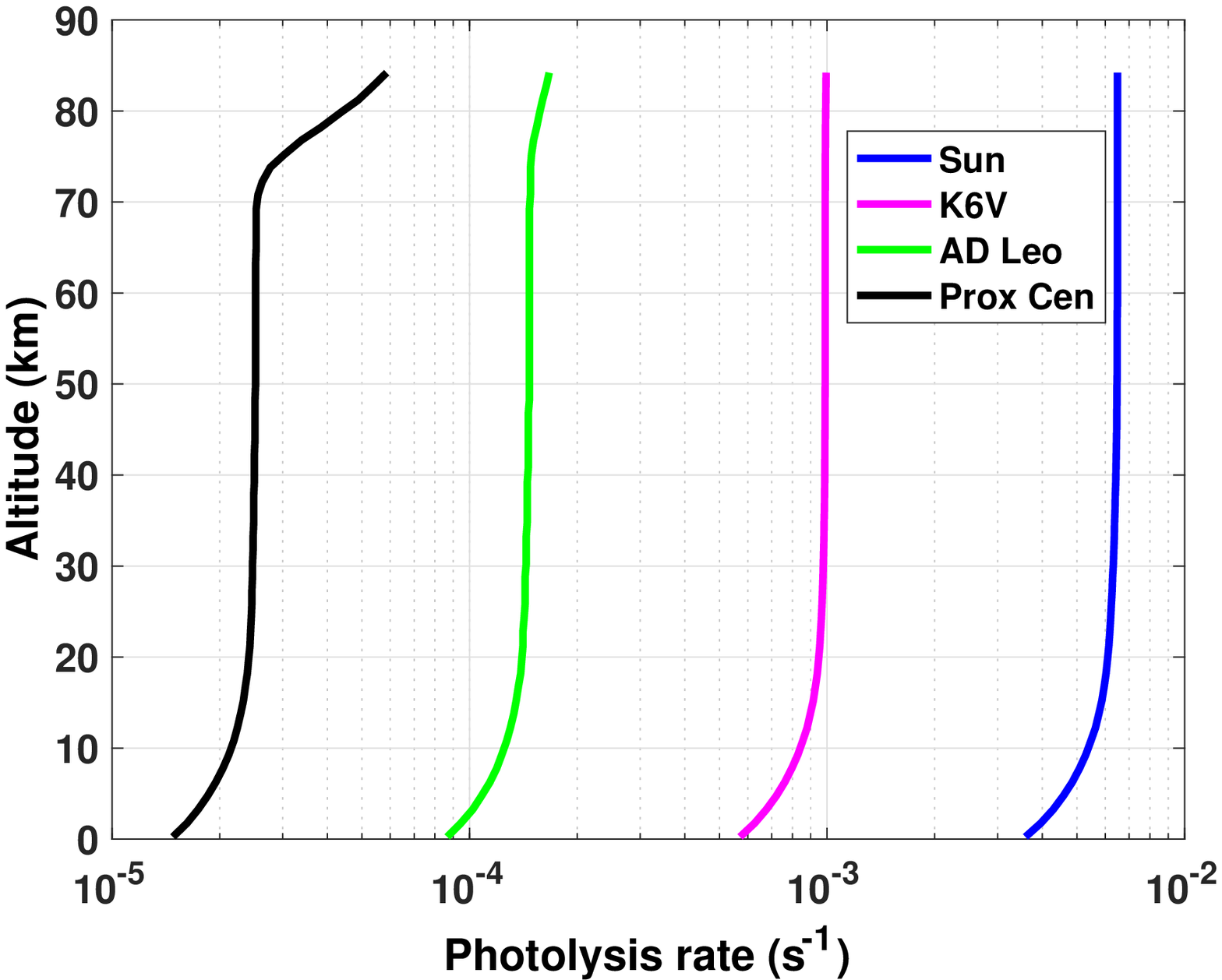}{0.49\textwidth}{(b)}
          }
    \caption{ (a): Mixing ratio profiles of O$_{3}$ around stars of different spectral types   on an Earth-like planet with 1x  (solid) and  20x (dashed) present Earth fluxes of NO$_{2}$. Below the troposphere ($\sim< 20$km), O$_{3}$ concentration rapidly increases around a Sun-like star (blue) compared to other star types. Because O$_{3}$ is a dominant production mechanism for NO$_{2}$ (See Table 1), the concentration of NO$_{2}$ ideally should increase. (b) However, as shown in this panel, the photolysis rate of NO$_{2}$ increases from the ground to up to 10km - 20km for all star types, as photons of wavelength between $0.29 - 0.42 \mu$m  penetrate into the troposphere. Consequently NO$_{2}$ mixing ratio {\it decreases} between $\sim 10 - 20$km (See Fig. \ref{fig:vmr}). Higher than 20km altitude, O$_{3}$ dominates the photolysis rate (because NO$_{2}$ photolysis is not increasing anymore), and as a result, NO$_{2}$ mixing ratio increases as well.
    } 
    \label{fig:photo}
\end{figure}

However, photolysis is not the only important factor . As shown in Table \ref{table1} and Fig. \ref{fig:photo}, O$_{3}$ also plays a major role in determining NO$_{2}$ concentration. Panel (a) in Fig. \ref{fig:photo} shows the O$_{3}$ mixing ratio profiles for various stars. For the Sun (blue), the O$_{3}$ concentration increases rapidly below $\sim 20$ km compared to other stars. O$_{3}$ participates in the dominant production reaction for NO$_{2}$, with the help of NO as shown in Table \ref{table1}. Ideally, this should increase the concentration of NO$_{2}$ below 20 km. However, as shown in panel (b) of Fig. \ref{fig:photo}, photolysis of NO$_{2}$ due to photons of wavelengths between $0.29 - 0.42\mu$m that penetrate into the troposphere dominate the destruction of NO$_{2}$, decreasing its mixing ratio (See Fig. \ref{fig:vmr}, panel a). While the photolysis rates generally increase for all stars below 20 km as shown in panel b of Fig. \ref{fig:photo}, it is the rate at which O$_{3}$ increases below this altitude that determines the slope  of decrease in NO$_{2}$ in the troposphere for planets around different stars. While the rapid increase in O$_{3}$ is mostly negated by the rapid photolysis and consequent decrease of NO$_{2}$ below $\sim 20$km for Sun-like stars (blue solid in Figs. \ref{fig:vmr} and \ref{fig:photo}), for other stars the O$_{3}$ concentration increases only a little from the surface to the tropopause (black,  green and magenta curves in Fig. \ref{fig:photo}). Consequently, the increasing slope of photolysis rate of NO$_{2}$ below $\sim 20$ km for these other stars slightly dominates (panel b), resulting in a  {\it larger} decrease in the mixing ratio of NO$_{2}$ compared to a Sun-like star (panel a in Fig. \ref{fig:vmr}). 

As a result, the column integrated NO$_2$ abundance increases moving from hotter stars to cooler stars (Table \ref{table2}).  The absorption cross sections for NO$_{2}$ and other key gases are shown in Fig. \ref{fig:opacity}. A Sun-like star produces more photons at wavelengths where NO$_{2}$ is photolyzed (between $290 - 420$nm), so the photolysis rates of NO$_{2}$ is higher for the planet around the Sun than for a planet around a cooler star (Table \ref{table1}). 

The result of all the dominant production and destruction reactions discussed above is that NO  and NO$_{2}$  decrease with altitude until  $\sim 20$\,km, into the stratosphere. In the stratosphere, ozone can generate NO$_2$ with reactions with NO; ozone's overlapping UV cross section with NO$_2$ also provides some UV shielding. At higher altitudes, above the ozone layer, photochemical processes, including NO$_2$ photolysis and reaction with OH radicals, draw down abundances. These reactions occur most markedly for the planet orbiting the Sun; NO$_2$ photolysis proceeds 1-2 orders of magnitude faster around the Sun compared to around the cooler stars. Reaction of NO$_2$ with OH to form HNO$_3$ occurs two orders of magnitude more efficiently around the Sun compared to the K6V star, and fully 4-6 orders of magnitude more efficiently around the Sun compared to the M dwarfs.

\begin{table}
\begin{center}
\begin{tabular}{ |p{4cm}|p{3cm}|p{3cm}|p{3cm}|p{3cm}|  } 

\hline
\centering
Reaction& Integrated reaction/photolysis rate for Sun (s$^{-1}$) & Integrated reaction/photolysis rate for K-dwarf (s$^{-1}$ & Integrated reaction/photolysis rate for 3390K star (s$^{-1}$) & Integrated reaction/photolysis rate for 3000K star (s$^{-1}$)\\ 
\hline
$ \mathrm{ CH_{3}O_{2}} + \mathrm{NO} \rightarrow \mathrm{CH_{3}O + NO_{2}} $ & 8.199 $\times 10^{10}$ & $8.623 \times 10^{10}$ & $6.850 \times 10^{10}$ & $1.139 \times 10^{10}$\\ 
\hline
 $ \mathrm{NO} + \mathrm{O_{3}} \rightarrow \mathrm{NO_{2} + O_{2}} $ & 2.919 $\times 10^{13}$ & $6.135\times 10^{12}$ & $2.054 \times 10^{12}$&$3.868 \times 10^{11}$ \\ 
\hline
$\mathrm{NO + O + M \rightarrow NO_{2} + M}$&$7.393 \times 10^{9}$& $3.248 \times 10^{8}$ & $4.515 \times 10^{7}$& $1.31 \times 10^{6}$\\
\hline
$\mathrm{NO + HO_{2} \rightarrow NO_{2} + OH}$& $5.521 \times 10^{11}$& $3.1 \times 10^{11}$ &$1.368 \times 10^{11}$&$4.514 \times 10^{10}$\\
\hline
$\mathrm{NO + NO_{3} \rightarrow 2NO_{2} }$& $1.099 \times 10^{10}$& $1.387 \times 10^{10}$ & $6.227 \times 10^{10}$& $3.152 \times 10^{10}$\\
\hline
$\mathrm{ HO_{2}NO_{2} + M}$ $\rightarrow$ $\mathrm{HO_{2} + NO_{2} + M }$& $1.713 \times 10^{11}$& $1.012 \times 10^{12}$ & $1.170 \times 10^{12}$&$7.157 \times 10^{11}$\\
\hline
$\mathrm{NO_{3} + h\nu \rightarrow NO_{2} + O }$& $2.949 \times 10^{10}$& $8.85 \times 10^{10}$ & $1.150 \times 10^{11}$& $2.974 \times 10^{10}$\\
\hline
$\mathrm{ NO_{2} + O\rightarrow NO + O_{2}}$ & $2.592 \times 10^{12}$ &$1.0 \times 10^{11}$ & $2.495 \times 10^{11}$& $2.857 \times 10^{10}$\\ 
\hline
$\mathrm{ NO_{2} + OH + M}$$\rightarrow$ $\mathrm{HNO_{3} + M}$ & $2.512 \times 10^{12}$ &$1.04 \times 10^{10}$ & $1.764 \times 10^{8}$&$4.973 \times 10^{6}$ \\ 
\hline
$\mathrm{O + NO_{2}\rightarrow  NO_{3}}$ & $1.219 \times 10^{10}$ &$7.234 \times 10^{8}$ & $5.338 \times 10^{8}$& $3.643 \times 10^{7}$\\ 
\hline
$\mathrm{O_{3} + NO_{2}\rightarrow  NO_{3} + O_{2}}$ & $2.658 \times 10^{10}$ &$1.169 \times 10^{11}$ &$2.014 \times 10^{11}$ & $7.278 \times 10^{10}$\\ 
\hline
$\mathrm{HO_{2} + NO_{2} + M} \rightarrow$ $\mathrm{HO_{2}NO_{2} + M}$& $1.78 \times 10^{11}$& $1.017 \times 10^{12}$ & $1.173 \times 10^{12}$&$7.168 \times 10^{11}$\\
\hline
$\mathrm{NO_{2} + h\nu \rightarrow NO + O}$& $2.724 \times 10^{13}$& $6.43 \times 10^{12}$ & $2.05 \times 10^{12}$& $4.318 \times 10^{11}$\\
\hline
%$\mathrm{O_{2} + h\nu \rightarrow O + O^{1}D}$& $2.986 \times 10^{11}$& $3.777 \times 10^{11}$ & &\\
%\hline
%$\mathrm{O_{2} + h\nu \rightarrow O + O}$& $6.125 \times 10^{12}$&$4.528 \times 10^{11}$ & &\\
%\hline
%$\mathrm{O_{3} + h\nu \rightarrow O_{2} + O^{1}D}$&$6.724 \times 10^{14}$& $2.149 \times 10^{13}$ & &\\
%\hline
%$\mathrm{O_{3} + h\nu \rightarrow O_{2} + O}$&$1.591 \times 10^{15}$& $2.956 \times 10^{14}$ & &\\
%\hline
%$\mathrm{O_{3} + h\nu \rightarrow O + O + O}$&$1.758 \times 10^{11}$&$3.07 \times 10{^9}$ & &\\
%\hline
\end{tabular}
\caption{Reactions that act as dominant sources and sinks for NO$_{2}$ (first column), and column integrated reaction or photolysis rates for an Earth-like planets around Sun (second column), K-dwarf (third column), 3390K star (fourth column) and 3000K star (fifth column). Bold font are production mechanism for NO$_{2}$, and normal font are loss mechanisms. The dominant sink is NO$_{2}$ photolysis and the dominant production mechanism is NO reaction with O$_{3}$ (in addition to the surface flux).}
\label{table1}
\end{center}
\end{table}

\begin{table}
\begin{center}
\begin{tabular}{ |p{3cm}|p{3cm}| p{3cm}|p{3cm}|p{3cm}|} 
\hline
\centering
Species & Sun & K6V (4715K) & AD Leo (3390K) & Proxima (3000K)\\ 
&(molecules/cm$^{2}$)&(molecules/cm$^{2}$)&(molecules/cm$^{2}$)&(molecules/cm$^{2}$)\\
\hline
\centering
NO$_{2}$ & $4.644 \times 10^{10}$ & $8.589 \times 10^{10}$ & $2.040 \times 10^{11}$ & $2.453 \times 10^{11}$\\ 
\hline
\centering
O$_{3}$ & $5.915 \times 10^{13}$ & $2.302 \times 10^{13}$ & $3.217 \times 10^{13}$ & $7.957 \times 10^{12}$\\ 
\hline
%$\mathrm{NO_{2} + h\nu \rightarrow NO + O}$& $2.724 \times 10^{13}$& $6.43 \times 10^{12}$ & $2.05 \times 10^{12}$& $4.318 \times 10^{11}$\\
%\hline
\end{tabular}
\caption{Column integrated number densities of NO$_{2}$ and O$_{3}$ (i.e, total number of molecules per unit volume of integrated along a column of atmosphere) on an Earth-like planet with 1x NO$_2$ flux around stars of different stellar spectral types. NO$_{2}$ is more abundant on a planet around cooler stars than around a Sun-like star, despite having more O$_{3}$ which is the dominant molecule in producing NO$_{2}$, because short wavelength photons are available more around a Sun-like star than a K or M-dwarf star. This results in higher photolysis rate  (destruction) of NO$_{2}$ around a Sun-like star (see Table \ref{table1}) reducing it's abundance .} 
\label{table2}
\end{center}
\end{table}

It is important to note that placing constraints on a planet's NO$_{2}$ abundance from its spectrum would not definitively answer whether the NO$_{2}$ is biologically or abiotically produced. One would need to estimate the production rates required to produce the observed NO$_{2}$ abundance and evaluate whether abiotic sources alone can sustain the inferred production rate.  
   
\section{Detectability of Nitrogen Dioxide}\label{sec:detectability}

\begin{figure}
    \gridline{\fig{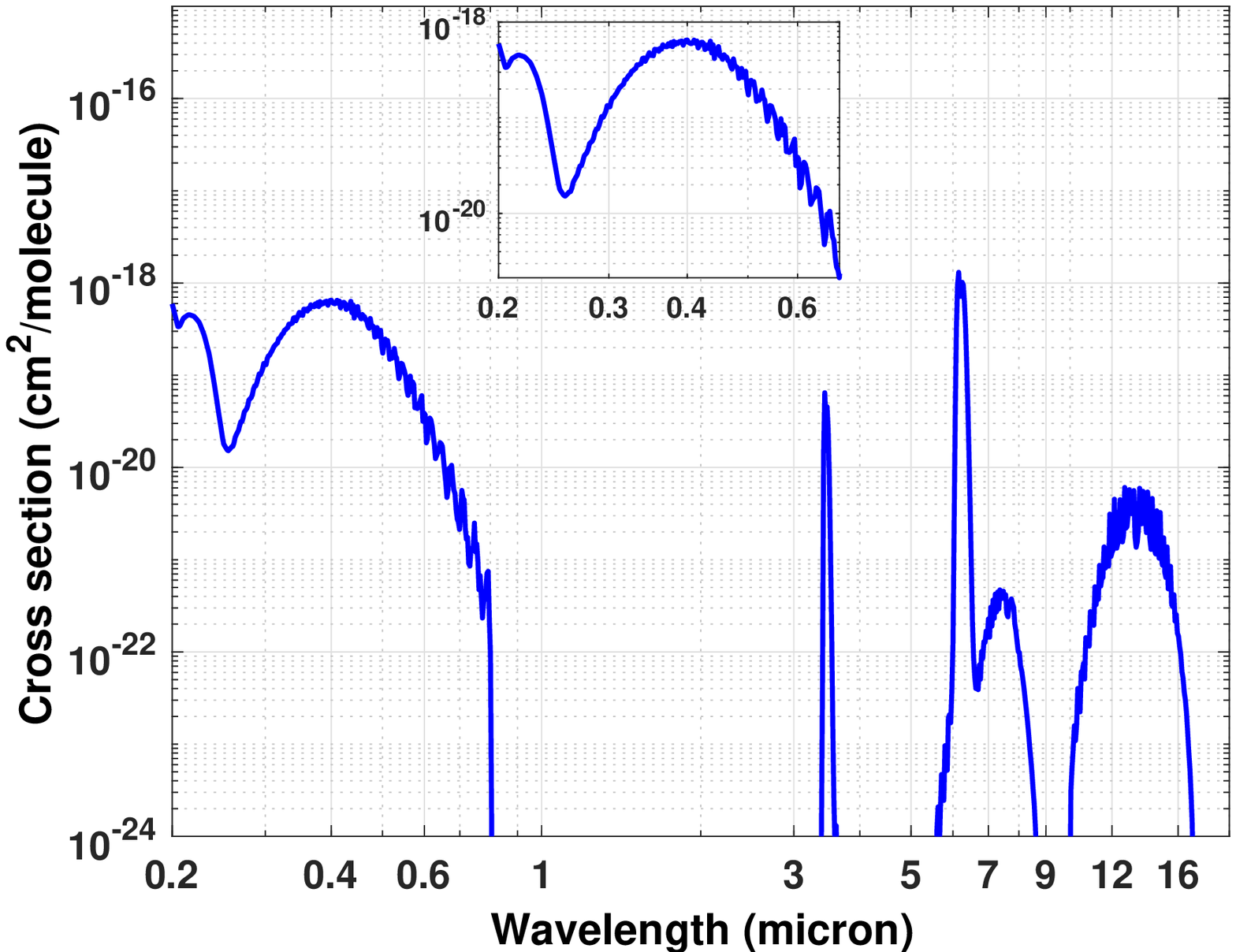}{0.49\textwidth}{(a)}
          \fig{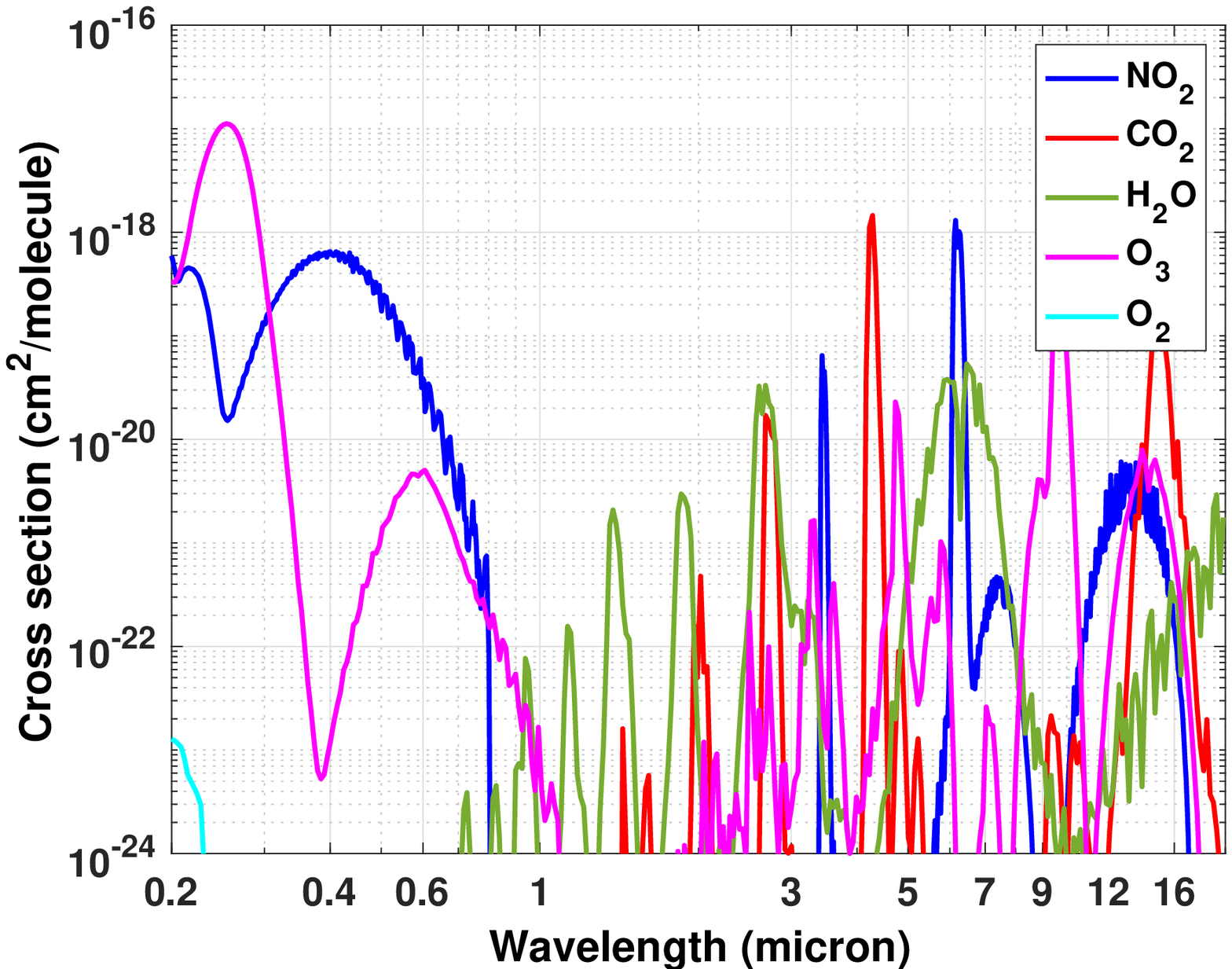}{0.49\textwidth}{(b)}
          }
    \caption{NO$_{2}$ absorption cross section as a function of wavelength (panel a). The broad absorption between 0.25-0.6\,$\mu$m is the dominant feature, and few other molecules absorb here. The inset figure focuses on the cross section in this wavelength region. Other features in the IR region ($\sim 3.5\,\mu$m, $6.4\,\mu$m and $10-16\,\mu$m) are relatively weaker and overlap with absorption from other gas species, in particular H$_{2}$O and CO$_{2}$ (panel b). } 
    \label{fig:opacity}
\end{figure}

The absorption cross section of NO$_{2}$ shows a broad absorption between 0.25-0.6\,$\mu$m, which has little overlap with absorption from other terrestrial molecular atmospheric constituents (Fig. \ref{fig:opacity}, panel b).  The main possible confusion would be related to aerosols with sub-micron sizes ($\sim 0.5\,\mu$m), which have absorption features that could mimic the exact same shape as NO$_{2}$. Considering the broad nature of the NO$_{2}$ spectral feature, a unique spectroscopic identification will be therefore ultimately challenging, and this investigation solely explores the hypothetical requirements for a possible detection for an absorption due to NO$_{2}$. Other absorption features are also present at $\sim 3.5\,\mu$m, $6.4\,\mu$m and $10-16\,\mu$m, but these overlap with absorption bands from H$_{2}$O, CO$_{2}$, and other species (Fig. \ref{fig:opacity}, panel b). In order to assess the detectability of NO$_{2}$ as a technosignature, we use the mixing ratio profiles from the 1-D photochemical model as input to the Planetary Spectrum Generator (PSG\footnote{https://psg.gsfc.nasa.gov/index.php}, \citet{psg:2018}) to simulate reflected light, and transit spectra. We estimate the signal-to-noise (SNR) of detecting NO$_{2}$ features. 
%The required observation times to observe transit spectra for plants around these G and K dwarfs were deemed infeasible, with 1000s of hours of observation time. 
PSG is an online radiative transfer suite that integrates the latest radiative transfer methods and spectroscopic parameterizations, and includes a realistic treatment of multiple scattering in layer-by-layer spherical geometry. It can synthesize planetary spectra (atmospheres and surfaces) for a broad range of wavelengths for any given observatory. 

\begin{figure*}
    \gridline{\fig{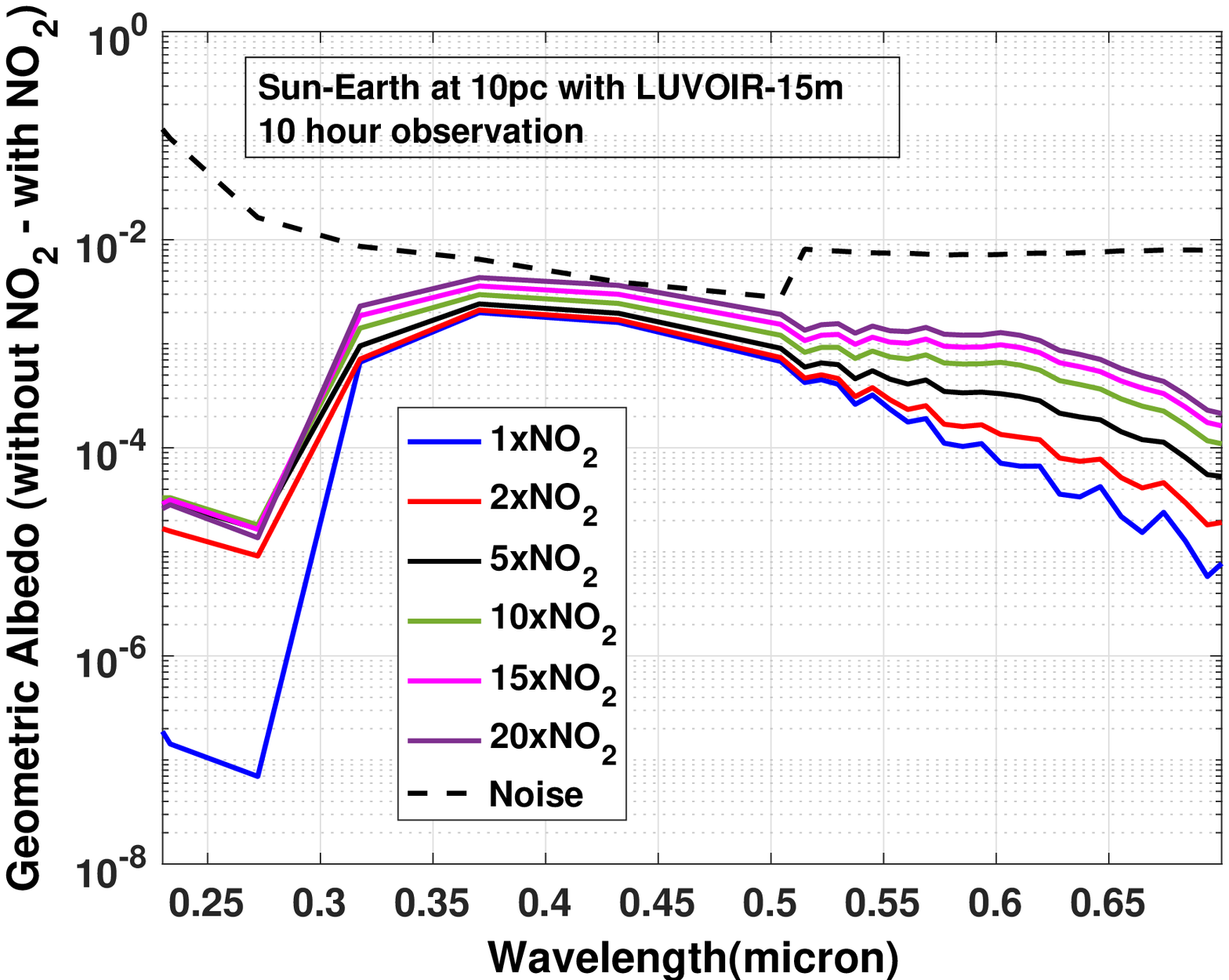}{0.49\textwidth}{(a)}
          \fig{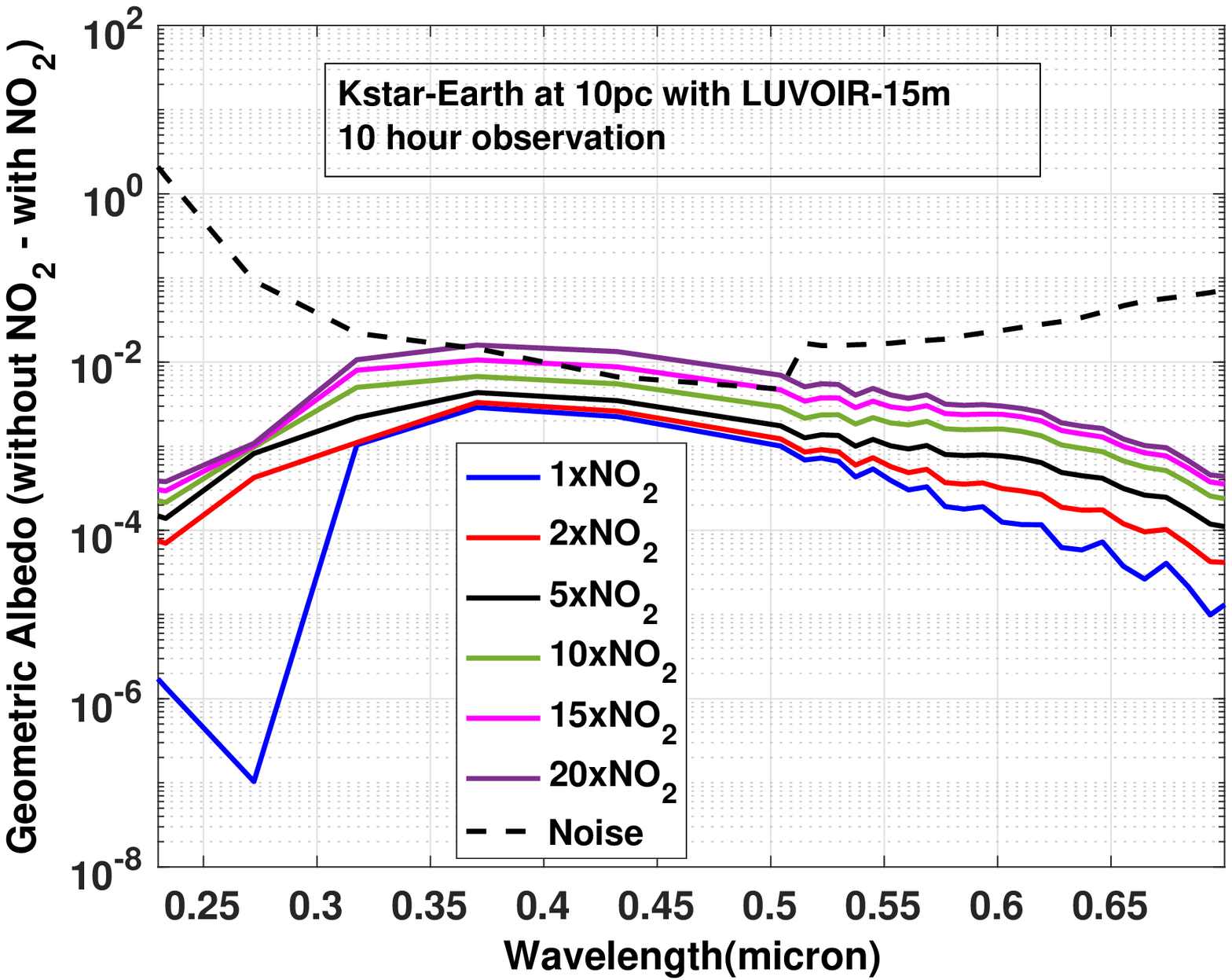}{0.49\textwidth}{(b)}
          }
    \caption{Geometric Albedo difference with and without NO$_{2}$ for an Earth-like planet around a Sun-like star (panel a) and around a K6V stellar spectral type (panel b) located at 10\,pc with varying NO$_{2}$ concentrations, assuming LUVOIR-A (15\,m) observing time of 10 hours. 1$\sigma$ noise model is also shown (dashed black).  The multiple factors in the legend are compared to the concentrations of present Earth level of NO$_{2}$ flux (8.64 $\times 10^{9}$ molecules/cm$^{2}$/s) implemented in our photochemical model of an Earth-like planet around each stellar type. 
    %The NO$_{2}$ features are more pronounced around the K6V star because the UV photons that destroy NO$_{2}$ are low around this stellar spectral type due to its red-shifted spectrum. 
    These are cloud free model results.} 
    \label{fig:spectra}
\end{figure*}

\begin{figure*}
    \gridline{\fig{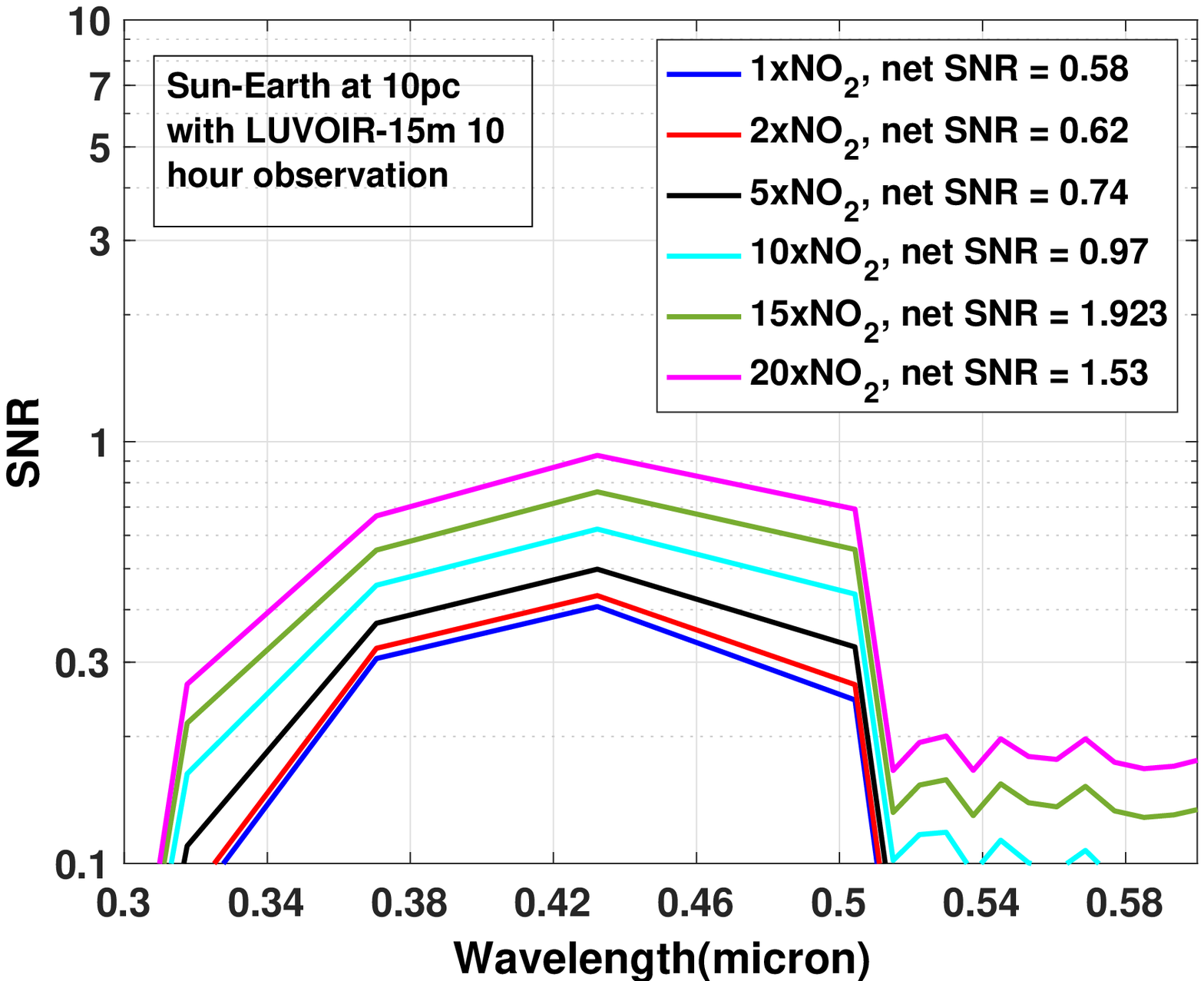}{0.49\textwidth}{(a)}
          \fig{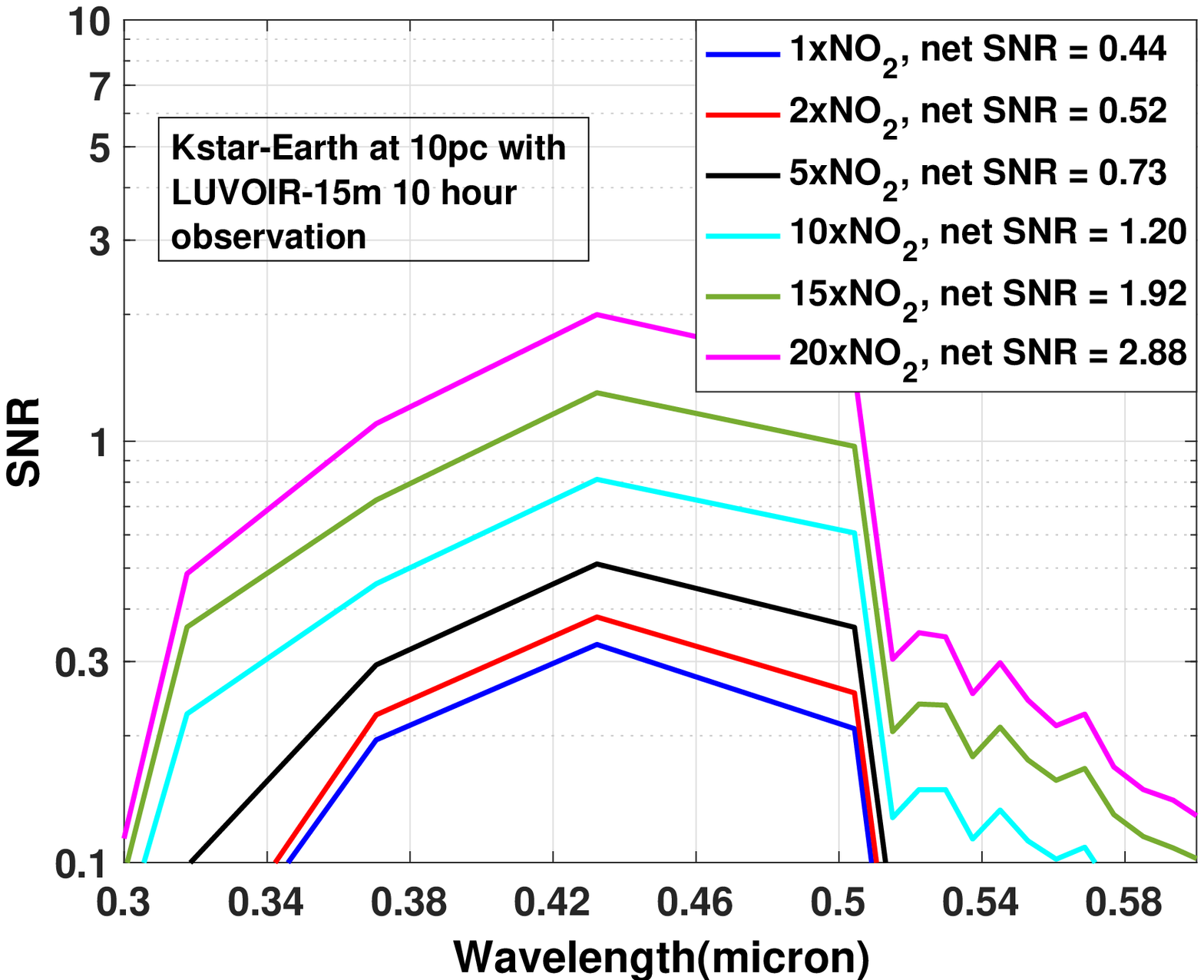}{0.49\textwidth}{(b)}
          }
    \caption{Calculated SNR values to detect various levels of NO$_{2}$ concentrations as a function of wavelength, around a Sun-like star (panel a) and for a K6V spectral type star (panel b) located at 10\,pc. The calculation assumed a LUVOIR-A (15m) type telescope with 10 hour observation time. While these plots show that at any given wavelength, NO$_{2}$ of any concentration is detected comparatively at a higher SNR around a K6V star than a Sun-like star, it will still be challenging to detect the feature within 10 hours. The NO$_{2}$ concentrations are generally higher around the K-dwarf star compared to an Earth-like planet around a Sun-like star, giving rise to this marginal increment in SNR around a K-dwarf star. These are cloud free model results.  }
    \label{fig:snr}
\end{figure*}

\begin{figure*}
    \gridline{\fig{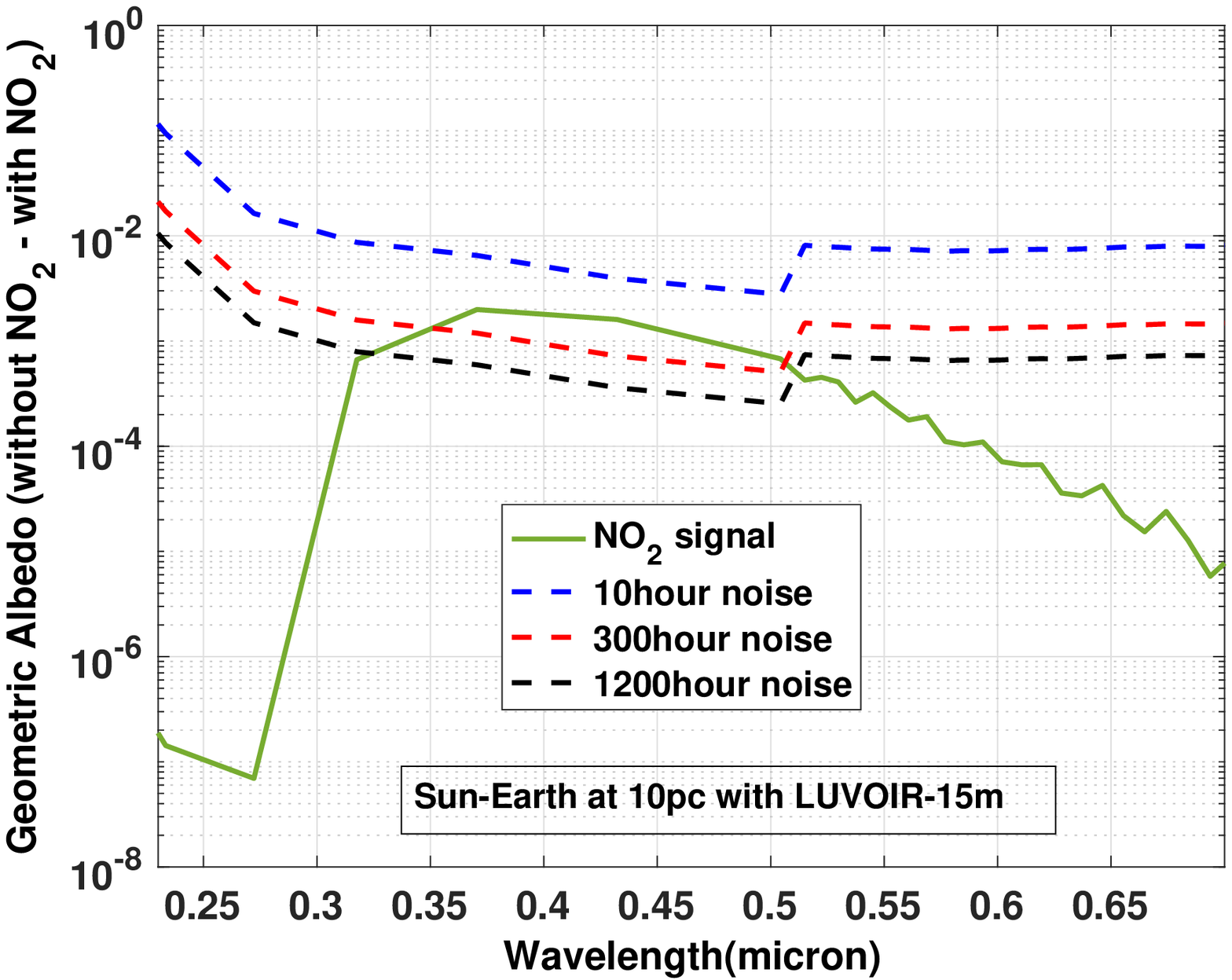}{0.49\textwidth}{(a)}
%          \fig{ELT.eps}{0.49\textwidth}{(b)}
          \fig{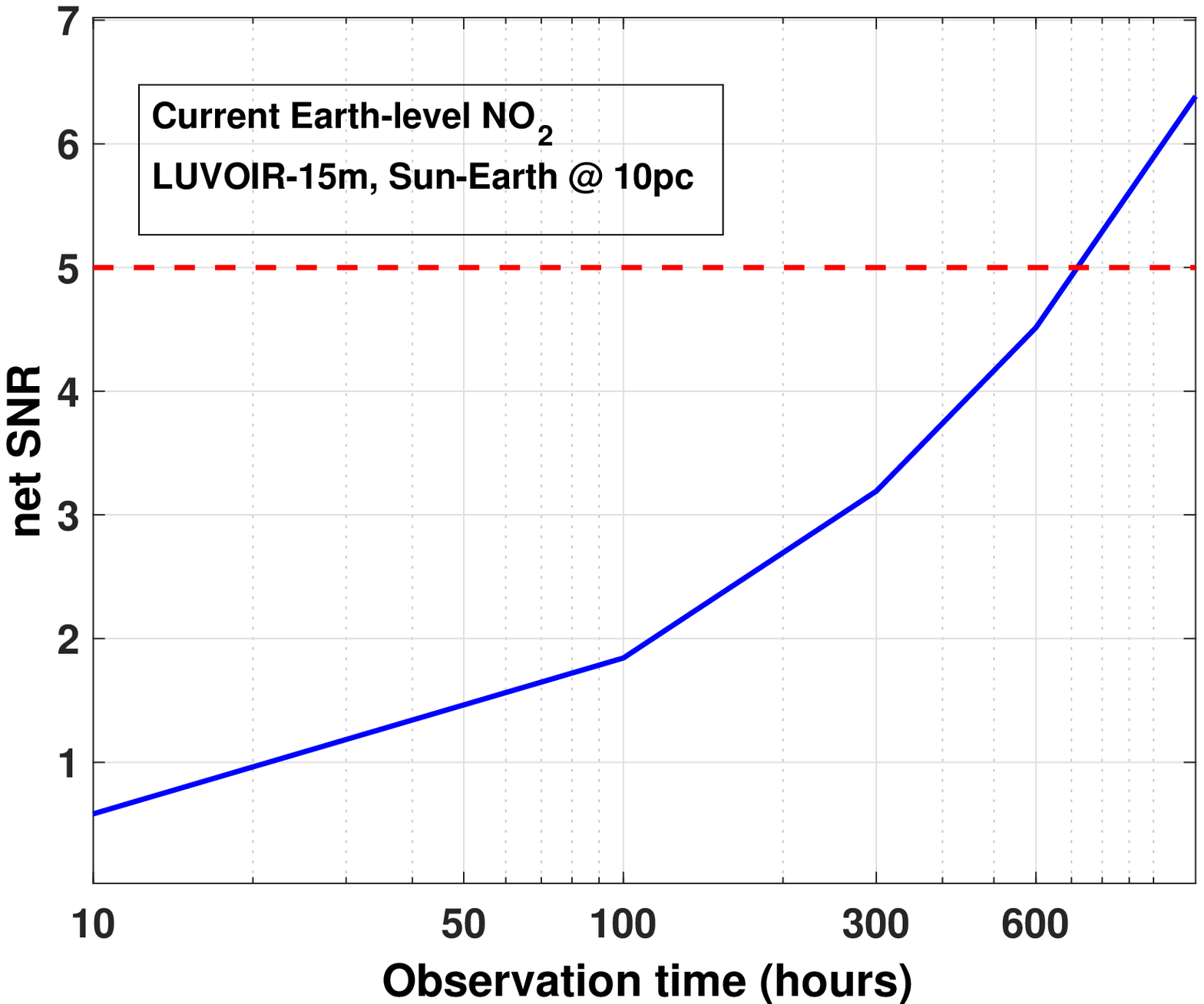}{0.49\textwidth}{(b)}
          }
    \caption{ (a) Geometric albedo difference with and without NO$_{2}$ as a function of wavelength for different observation times with LUVOIR-15m telescope to detect a present Earth-level NO$_{2}$ amount on a Sun-Earth system at 10pc. (b) Integrated SNR (over the wavelengths) versus the amount of observation time needed for the same system configuration. For example, to detect an Earth level NO$_{2}$ with SNR $\sim 5$ (red-dashed line), LUVOIR-15m would need $\sim 400$ hours of observation time. For comparison, Hubble's large programs such as the Ultra Deep Field (UDF) and CANDLES surveys used between $\sim 400 - 900$ hours of observation time over a period of 1-3 years. See text for more details.}
    \label{fig:elt}
\end{figure*}

We performed simulations with PSG to generate reflected light spectra (Fig. \ref{fig:spectra}) of planets around Sun-like star and a K-dwarf star. We then calculated required SNR to detect the NO$_{2}$ feature (Fig. \ref{fig:snr}) between 0.2- 0.7 $\mu$m. For these simulations, we assumed a LUVOIR-A like telescope (15 meter) observing with the ECLIPS (Extreme Coronagraph for LIving Planetary Systems).\footnote{https://www.luvoirtelescope.org/, section 1.11.2, page 75} This instrument is an internal coronagraph with the key goal of direct exoplanet
observations. It has three channels: NUV (0.2–0.525 $\mu$m), visible
(0.515–1.030 $\mu$m) and NIR (1.0–2.0 $\mu$m).  The NUV channel is capable of high-contrast
imaging only, with an effective spectral resolution of R$\sim 7$. The optical
channel contains an imaging camera and integral field spectrograph (IFS) with R=140. For our spectral simulations, we use NUV (R=6) and visible (R=70) channels, as the NO$_{2}$ cross section spans from UV into visible wavelengths (See Fig. \ref{fig:opacity}). Because the NO$_{2}$ feature is quite broad in the NUV to visible region, a low resolution of R=6 and R=70 is suffice to resolve the feature, at the same time maximizing the SNR. We calculated wavelength dependent SNR shown in Figs. \ref{fig:spectra}, \ref{fig:snr}, \ref{fig:elt}, \ref{fig:clouds} and \ref{jwst-ost} as the difference between the spectra with and without the NO$_{2}$ feature, divided by the noise simulated by PSG for the instrument under consideration (see section 5.3 of \citet{psg:2018}, and also the PSG website\footnote{https://psg.gsfc.nasa.gov/helpmodel.php$\#$noise} where the noise model is discussed in detail).  The ``net SNR'' is calculated by summing the squares of the individual SNRs at each wavelength within a given band (either NUV or VIS), and then taking the square root. This methodology  is largely insensitive to SNR, as long as the feature is resolved by the spectrum. See also Appendix \ref{appendix:cg} for a comparison between the PSG coronagraph noise model and a complementary noise model \citep{Robinson2016, Lustig-Yaeger2019cg}, showing highly comparable results for the photon count rates and resulting spectral precision. We considered the planets around both the Sun-like and K6V stars to be located at 10\,pc distance, residing in the respective habitable zones (HZs) of their host stars as calculated from \citet{Kopparapu2013, Kopparapu2014}, and observed at a phase angle of $45^{\circ}$ ($0^{\circ}$ is secondary eclipse, and $180^{\circ}$ is transit). For this feature to be detected, the planet need not be in the HZ, as will be explained later in the discussion section (\ref{sec:discussion}). 

In both the panels of Fig. \ref{fig:spectra}, the difference of the geometric albedo spectrum with and without NO$_{2}$ are shown for different levels scaled by factors of current Earth levels in a 10 hour observation with LUVOIR-15m telescope. The corresponding noise is shown as dashed curve. For the Sun-like star (panel a)  even very high (20x) concentrations of NO$_{2}$ compared to the present Earth levels barely reach the $1-\sigma$ noise level in the strongest wavelength region.  In panel b, increasing the nominal abundance to higher concentrations on a planet around a K-dwarf produces only a marginal improvement over a Sun-like star, with the highest concentration (20x) reaching just above the noise level. This is likely because the column number density of NO$_{2}$ on a planet around K-dwarf star seems marginally larger (Table \ref{table2}) due to less photolysis rate (last row, Table \ref{table1}).  As discussed above, the enhanced NO$_{2}$ absorption on the K-dwarf planet compared to the planet around the Sun-like star is driven by the photochemistry.

In Fig. \ref{fig:snr}, we show the calculated signal to noise ratio (SNR) values of the  features shown in Fig. \ref{fig:spectra} as a function of wavelength for 10 hour exposure times with a LUVOIR-A like telescope for wavelengths relevant to the NO$_{2}$ feature. The ``net SNR'' indicated in these figures is calculated by summing up the squares of the SNR from each wavelength band and then taking the square-root (see Eq.(6) of \cite{Jake2019}).
Fig. \ref{fig:snr}(a) shows that for planets around Sun-like stars even an increase of 10x in the NO$_{2}$ flux is not enough to detect the feature with any meaningful SNR within 10 hours of observation.  Any lower amount of NO$_{2}$ would need even more longer observation times.

Fig. \ref{fig:snr}(b) shows SNR as a function of the same wavelength range for a planet around K-dwarf star. The combined effects of more NO$_{2}$ and better planet-star contrast ratio relative to the planet orbiting the Sun (a K6V dwarf is only about one tenth as luminous as a G2V dwarf)  makes only a marginal difference in SNR that can be reached in the same time as Sun-like star. 

While these SNR may not look promising, there is an interesting question that one can ask and explore an answer: How much LUVOIR-15\,m time is needed to detect present Earth-level concentration of NO$_{2}$ around a Sun-like star at 10\,pc? Fig.\ref{fig:elt} (a) shows Geometric albedo spectrum difference with and without NO$_{2}$ as a function of wavelength for 
%this Star-planet configuration and 1x NO${2}$ level. To obtain a SNR of ~3, 5 and 7, it would take nearly
300, 600 and 1200 hours of LUVOIR-A time, respectively. Also shown in dashed lines are the corresponding noise levels for each of these observation times. The present Earth-level NO$_{2}$ seems to be well above the noise level after 300 hours of observation time (compare the solid green curve with red-dashed line) indicating that it might be detectable. To find out with what SNR it would be detectable, Fig. \ref{fig:elt} (b) shows the ``net SNR'' to detect present Earth-level NO$_{2}$ as a function of observation time. To achieve a net SNR of 5 (dashed red line), it would take LUVOIR-15\,m about 400 hours. 
For comparison, to obtain the Hubble Ultra Deep Field image, $\sim 400$ hours of actual observation time ($\sim 1$ year in real time) was needed \citep{beckwith2006hubble}. In fact, Hubble has done even larger programs such as the CANDLES galaxy evolution survey \citep{grogin2011} with 902 orbits ($\sim 900$ hours of observation time assuming $\sim 1$ hour per orbit). This took about 3 years in real time. However, these large programs also obtained data on a huge sample size with thousands of galaxies. LUVOIR is envisaged to be 100$\%$ community competed time and the final report of LUVOIR team laid out a DRM in which comparable allocations of time were spent on general astrophysics observations and exoplanet detection and characterization observations during the first 5 years of the mission. So, over the course of the  nominal LUVOIR mission lifetime of about 5 years, it may be possible to take data with $\sim 400$ hour observation time on a prime HZ planet candidate(s) within 10\,pc, to potentially obtain a SNR $\sim 5$ for a present Earth-level NO$_{2}$ feature on a Earth-Sun system at 10\,pc. An even more interesting aspect is that, we can place upper limits on the amount of NO$_{2}$ available on that planet as we spend more observation time on a prime HZ candidate. This could potentially indicate the presence or absence or the level of technological civilization on that planet.
%LUVOIR is planned to have ~4000 hours of observation time {\it per year} conservatively  for all science, and 40$\%$ ( $\sim$1600 hours {\it per year})  of it for exoplanet science. 

%{\bf We have performed another calculation to estimate the required SNR to detect the NO${_2}$ feature in direct imaging on Proxima Centauri b (the planet is not known to transit) for a 10 hour observation using a 30 meter ground-based telescope, such as the ELT. Interestingly, Fig. \ref{fig:clouds}(b) shows that a present-Earth NO$_{2}$ level, if present, can be detected with  a SNR of $> 5$ within 10 hours of observation. Higher NO$_{2}$ amounts can be detected with higher SNRs. The reason for this high SNR detection in a relatively short time is because Proxima Centauri is a M-dwarf star, and short wavelength photons that can destroy NO$_{2}$ are far less in number compared to Sun-like or K-dwarf stars (See Fig. \ref{fig:vmr} (b)). Consequently, NO$_{2}$ molecules have longer lifetime making them more abundant. We should note that this analysis needs to be done with a more realistic ELT-30\,m design specifications and noise models. However, the astrophysical fact that M-dwarfs have fewer short-wavelength photons, and the chemistry of planets around M-dwarf stars remains unchanged. Therefore, we anticipate that any future studies with realistic simulations may be close to our net SNR values.}

%between 15x to 20x present Earth level NO$_{2}$ is needed on Prox Cen b. }

\section{Discussion} \label{sec:discussion}

\begin{figure*}
\gridline{\fig{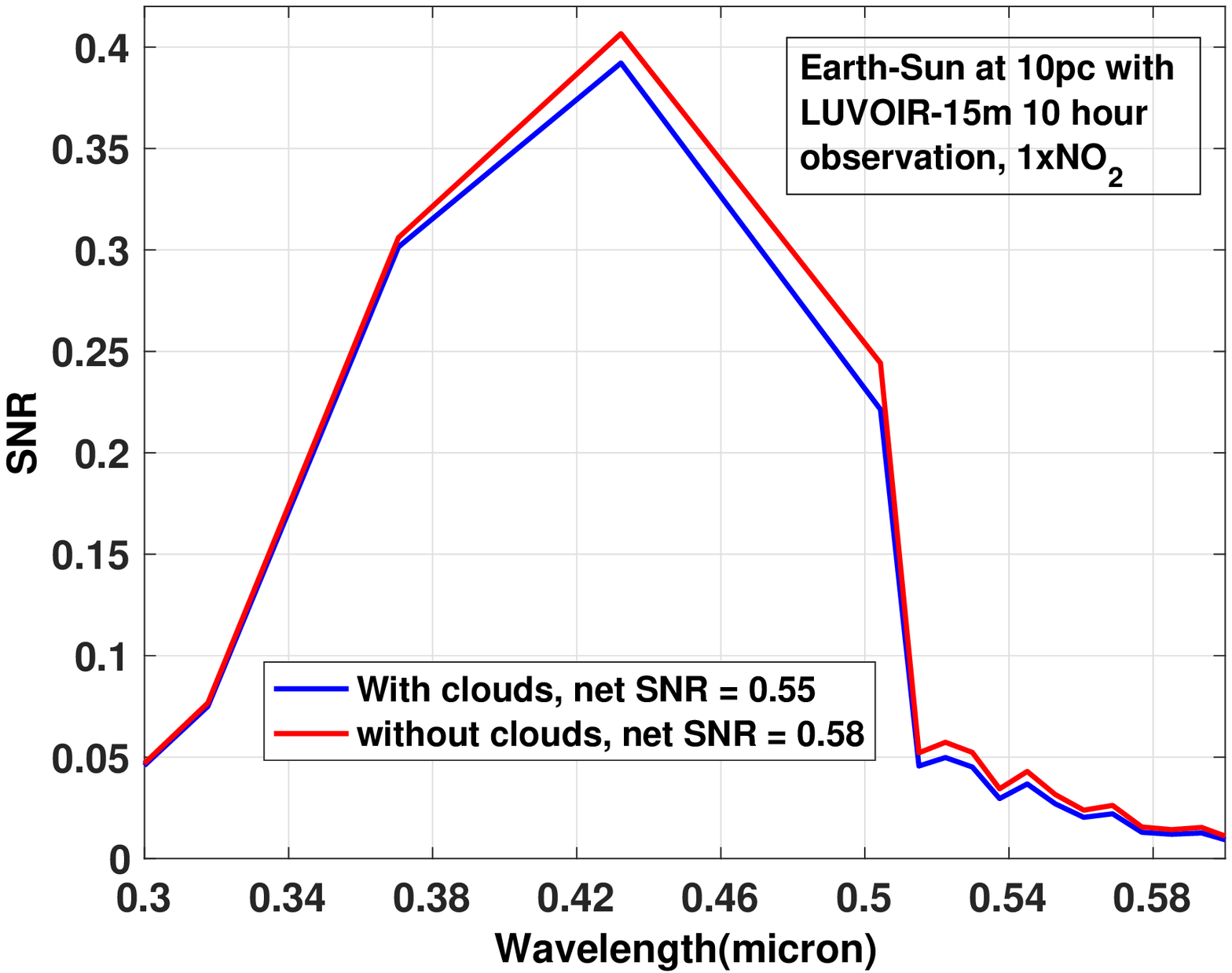}{0.49\textwidth}{(a)}
          \fig{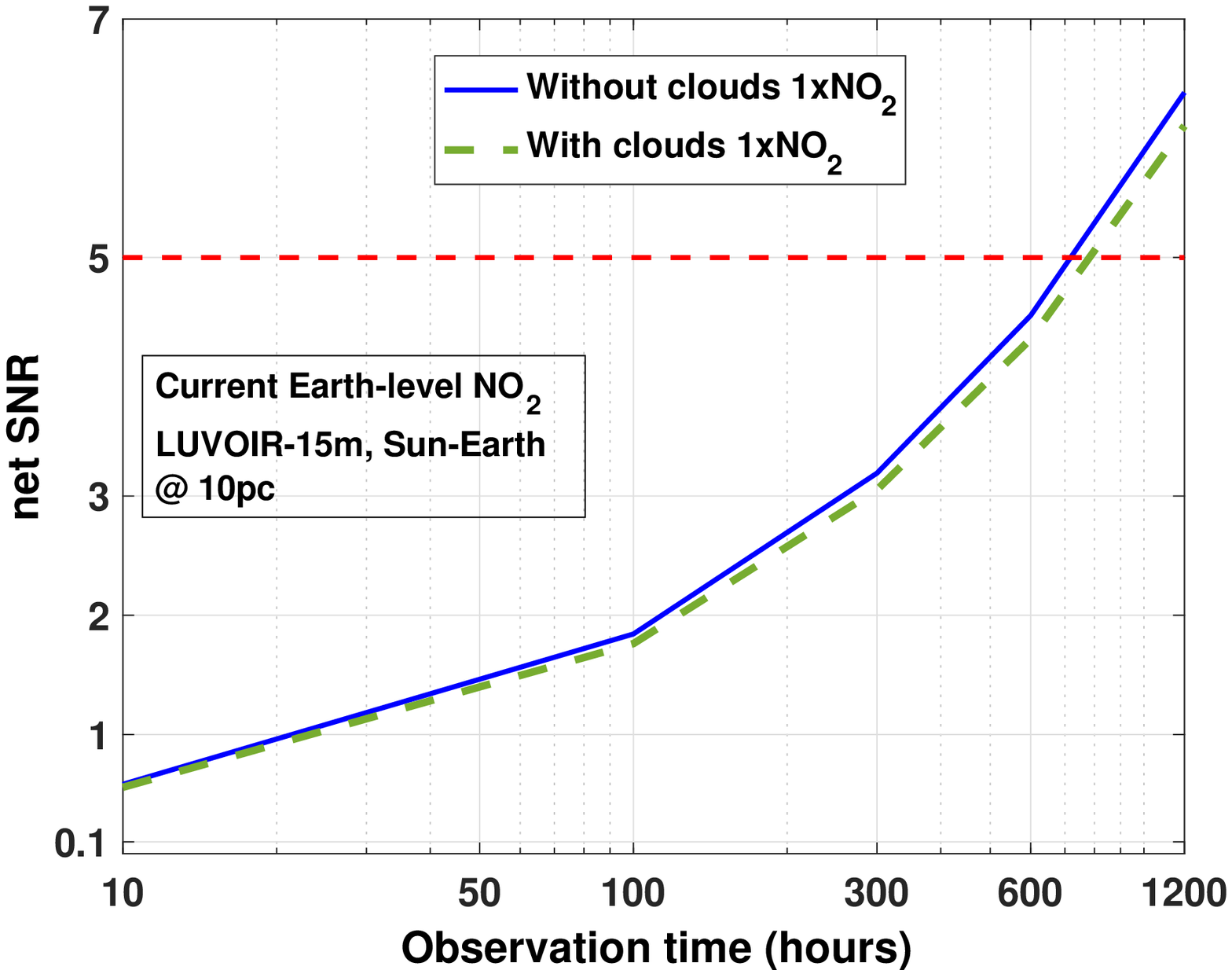}{0.49\textwidth}{(b)}
          }
\caption{Effect on the SNR of a geometric albedo spectrum with (blue solid)  and without (red solid) clouds on an Earth-like planet (1xNO$_{2}$) around  Sun-like star. Water clouds absorb in the same wavelength region as NO$_{2}$ absorption bands, thus reducing the SNR and potentially causing confusion source. %Comparing this figure with Fig.\ref{fig:snr}, the SNR is enhanced by several factors with clouds included.
%We used a 1-D climate and phothochemistry model to generate the Earth profile. However, these results indicate that one should really perform these calculations with a 3-D climate where clouds are simulated self-consistently. 
(b) Similar to Fig.\ref{fig:elt}b, integrated SNR (over the wavelengths) versus the amount of observation time with (green dashed) and without (blue solid) clouds. The time to reach a SNR = 5 is slightly longer with clouds.
%The required SNR versus wavelength for ground-based 30 meter class telescope direct imaging observations of Prox Cen b to detect different levels of NO$_{2}$. Only levels above 5x can achieve a SNR of $\sim 3$ in a 10 hour observation. 
}
\label{fig:clouds}
\end{figure*}

While the results from the previous section provide a preliminary study of NO$_{2}$ as a potential technosignature, some caveats need to be mentioned. 
First, we have performed 1-D photochemical model calculations using a modern Earth template generated from a 1-D radiative-convective, cloud-free climate model from \cite{Kopparapu2013}. Clouds can significantly effect the observed spectrum and potentially alter the calculated SNR. To test this, we have prescribed water-ice clouds (particle size $25 \mu$m) between 0.001 - 0.01bar, and liquid water clouds (particle size $14 \mu$m) between 0.01bar - 0.1bar in PSG.  Fig. \ref{fig:clouds} shows SNR as a function of wavelength for an Earth-like planet around a Sun-like star
at 10\,pc distance observed with the LUVOIR-15m telescope for 10 hours with (blue solid) and without (red solid) clouds. The absorption cross-sections of water and ice clouds are in the same wavelength region as the peak NO$_{2}$ absorption which further masks the NO$_{2}$ feature in this band. 
%(blue solid) 
%and around a K-dwarf star (red colour)
 %In contrast to the low values of SNRs shown in Fig. \ref{fig:snr}, here we find that clouds can significantly boost the NO$_{2}$ feature signal by several factors. Particularly, the SNR for a Sun-Earth system at 10pc becomes quite achievable with 10 hours of observation. %Note that, now, the SNR for a K-star seems {\it lower} than for the Sun-like star, likely arising from the availability of short-wavelength photons around a Sun-like star increasing the albedo of the planet.
%Essentially, a planet around a Sun-like star appears `brighter' boosting the signal, even though the NO$_{2}$ is more readily photochemically destroyed around a Sun-like star.
We should caution that this is all based on ad-hoc prescription of clouds at a certain height, and a more rigorous analysis using 3-D climate models which can simulate self-consistent and time-varying cloud cover need to be performed. We leave that for future study.  

%First, in our photochemical model, we increased the abundance of NO$_{2}$ by multiplying the base case (1x) mixing ratio profile values for a given stellar spectral type by factors of 2x, 5x, 10x, and 30x to obtain a first-order estimate showing how increasing NO$_{2}$ concentrations affect the spectrum. In future, to explore how different NO$_{2}$ production rates might impact the spectrum, one could adjust the NO$_{2}$ flux at the lower boundary of the photochemical model.  Second, this study, which focused on reflected light spectra, examined only the Sun and a K6V dwarf, {\bf because these stars are particularly amenable to direct observations due to the sufficient angular separation of the HZs of G \& K-dwarfs \citep{Arney2019}}. To examine NO$_{2}$ in a broader stellar context, additional spectral types should be studied in future. 
%We performed such calculations, and found that the geometric albedo spectrum of the planet around Sun-like star, shows little variation for 1x, 2x, 5x, 10x, and shows  significant change only for 30x the concentration of NO$_{2}$. However, for the planet around K-dwarf star,  is 
Secondly, we have used the 15\,m architecture of LUVOIR-A, and the SNR values we report are a best case scenario owing to its large mirror size. Other telescope architectures such as LUVOIR-B (8\,m) or HabEX\footnote{https://www.jpl.nasa.gov/habex/pdf/HabEx-Final-Report-Public-Release-LINKED-0924.pdf} (a 4\,m mirror accompanied by a coronagraph and a starshade) may need more observation time than shown in Fig. \ref{fig:snr} and Fig. \ref{fig:clouds} to detect NO$_{2}$ features. 

\begin{figure*}
    \gridline{\fig{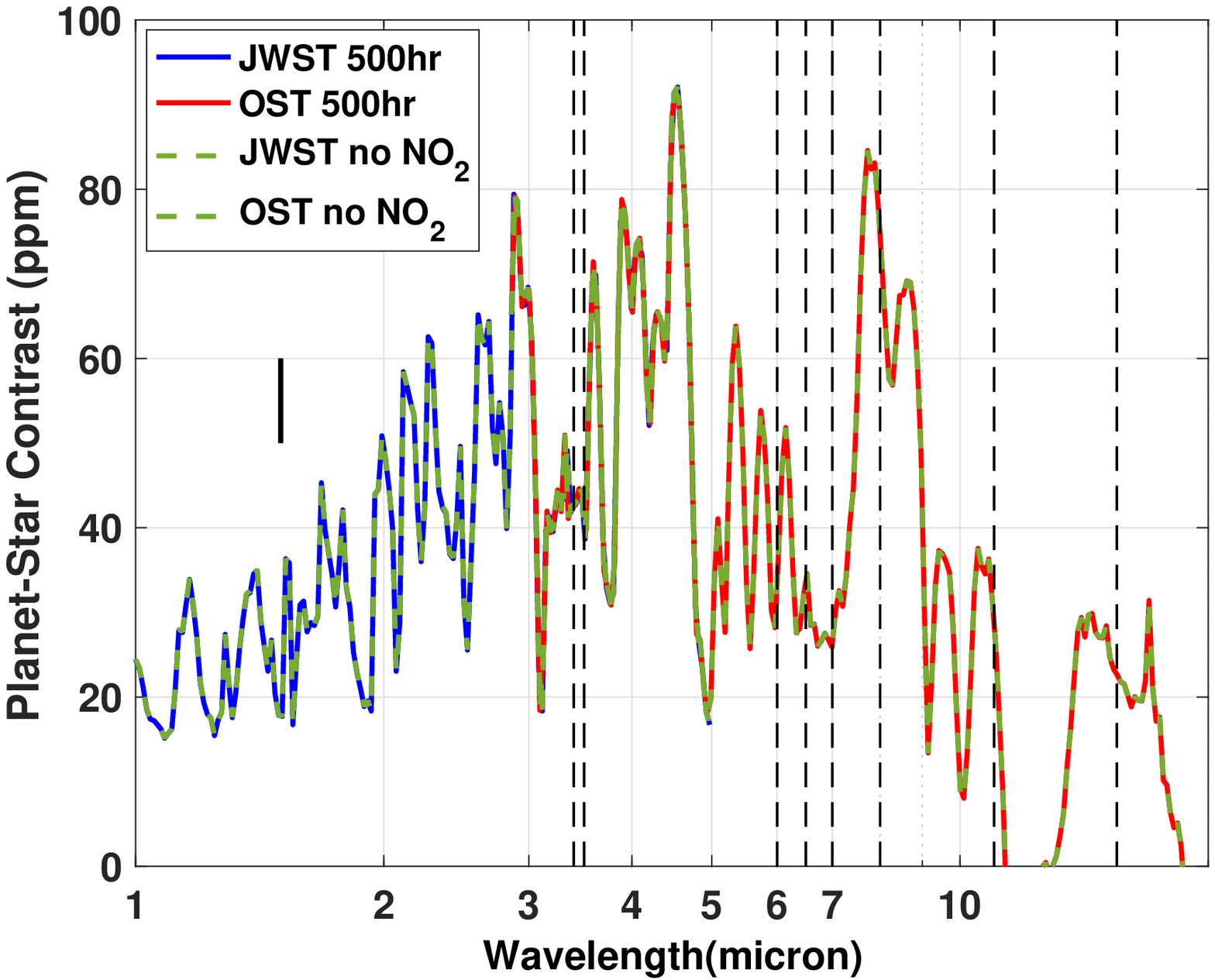}{0.49\textwidth}{(a)}
          \fig{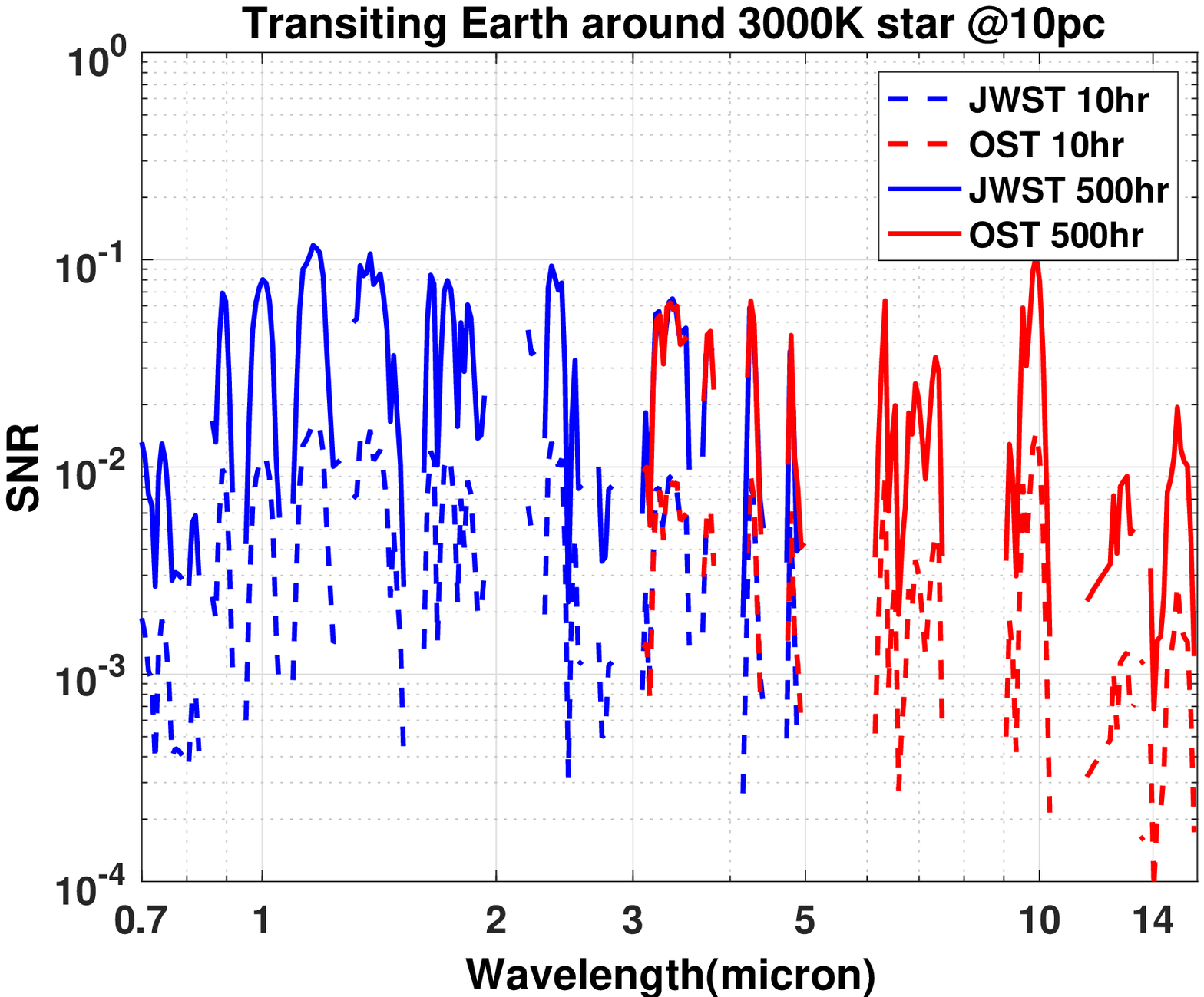}{0.49\textwidth}{(b)}
          }
    \caption{ (a) Transit spectrum of NO$_{2}$ in the near-IR and mid-IR region on a HZ Earth-like planet around Proxima Cen-like star ($T_{eff} = 3000$K) located at 10pc with 20x present Earth-level NO$_{2}$ fluxes using JWST NIRSpec (blue) and OST MISC (red) observations. No clouds were included. The vertical solid black line indicates the error bar, and dashed lines indicate NO$_{2}$ absorption bands in the IR (b) Both 10 hour (dashed) and 500 hour (solid) observations indicate it is very challenging to observe even 20x NO$_{2}$ abundance in transit observations in the IR because of several other overlapping gases in this region that have a stronger absorption than NO$_{2}$. See Fig. \ref{fig:opacity}(b).}
    \label{jwst-ost}
\end{figure*}

As shown in Fig. \ref{fig:opacity}, NO$_{2}$ also absorbs in the infrared (IR) part of the spectrum, particularly between 3.2-3.7\,$\mu$m, 5.2-8.9\,$\mu$m and 9.7-18\,$\mu$m. However, the absorption in these regions is either weak across the band compared to the 0.25-0.65\,$\mu$m visible band,  or limited to a very narrow region of the spectrum. Consequently, detecting NO$_{2}$ in transit spectroscopy with either JWST or the flagship mission concept study Origins Space Telescope (OST) would be challenging. Nevertheless, we tested this with PSG, and the results are shown in Fig. \ref{jwst-ost}. We placed a planet like Proxima Cen b around it's host star at 10\,pc assuming that it transits, with 20x NO$_{2}$ abundance to maximize the signal. We used JWST NIRSpec and OST MISC-Transit instrument for the $\sim 3\,\mu$m and $\sim 6\,\mu$m wavelength regions for the detection of NO$_{2}$. While OST has greater performance than JWST, the observations here are limited by masking features from H$_{2}$O and CO$_{2}$ in the near-IR. Even at 20x NO$_{2}$ from our photochemical model H$_{2}$O features completely dominate the $\sim 6\,\mu$m region of NO$_{2}$ absorption (Fig.\ref{jwst-ost}(a)). The SNR for a 10 hour and 500 hour observation times for both telescopes is shown in Fig.\ref{jwst-ost}(b). Even with large observation times, it would be very challenging to detect the NO$_{2}$ feature with any meaningful SNR.

%and found that it would take at least several hundred hours of JWST time with either NIRSpec or MIRI. The ground-based surveys are limited by sky background beyond 5\,$\mu$m, therefore the $\sim 3\,\mu$m band is the most accessible one. However, even then it would take hundreds of hours of observation time.  We also tested PSG with the Origins Space Telescope (OST) MISC-Transit instrument for the $\sim 3\,\mu$m and $\sim 6\,\mu$m wavelength regions for the detection of NO$_{2}$. While OST has greater performance than JWST, the observations here are limited by masking features from H$_{2}$O and CO$_{2}$ in the near-IR. Even at 30x NO$_{2}$ from our photochemical model for on an Earth-like planet in the HZ around a M-dwarf like Proxima Centauri,\footnote{While the actual planet in the HZ of Proxima Centauri, Prox Cen b, is not known to transit, we assumed a star {\it similar} in temperature and SED located at 10pc.} H$_{2}$O features completely dominate the $\sim 6\,\mu$m region of NO$_{2}$ absorption. We did not check OST results in thermal emission spectra, or for a planet with lower H$_{2}$O or CO$_{2}$. We leave these calculations, and specifically focusing on M-dwarf planets to a later study.

 A space-based nulling interferometer such as ESA's LIFE (Large Interferometer for Exoplanets) mission concept \citep{life2018,quanz2018} could potentially detect mid-IR ($5 - 20\mu$m) features in direct imaging spectra. While we are unable to assess quantitative limits on SNR at this time for this mission, we speculate that phase dependent thermal emission spectroscopy \citep{wolf2019, gabby2020} may be another way to detect NO$_{2}$ feature. 

 Historically, the United States NO$_{2}$ concentrations have varied (gone down) by a factor of 3 over a period of 40 years, from 1980-2019.\footnote{https://www.epa.gov/air-trends/nitrogen-dioxide-trends} Therefore, we can expand the possibilities of detecting a technological civilization at the stage where Earth civilization was 40 years ago. It is possible to imagine a more highly industrialized society that could possibly operate in the regime of $5\times$ Earth NO$_{2}$ level making it possible to detect it with LUVOIR-15m with even less observation time than for present-Earth conditions. We should stress here that when we mean a technological civilization, it does not necessarily mean a much more advanced society than current Earth level. Just like the search for biosignatures encompass `Earth-through time' with different stages of Earth's biosphere evolution, we could do a similar search for a `technosphere' at different stages of a technological civilization.
 
 It is possible that atmospheric technosignatures, in particular industrial pollutants like NO$_{2}$, are short-lived. However, this is comparable to searches for radio technosignatures where the transient nature of the radio communicative civilizations may also be short-lived. Furthermore, it may be that an industrialized society that is prone to emit NO$_{2}$ as a byproduct of their combustion technology may also have radio communication capabilities, just like us. In this respect, a search for radio technosignatures can be performed if NO$_{2}$ is detected on a potential habitable planet.

If we are looking for NO$_{2}$ as a technosignature, and not as a biosignature, then it may appear that one  need not limit the search to known planets in the HZ. A technological civilization can possibly inhabit even an adjacent barren planet (like Mars in our Solar System), and use the atmosphere as a waste dump of NO$_{2}$ emissions. Or they may prefer to live sub-surface on a HZ planet and release ``waste'' NO$_{2}$ into the atmosphere. Speculations are endless. However, industrial NO$_{2}$ on Earth is produced by essentially burning biomass (coal, petroleum products) that have been excavated to fuel the civilization (We note that NO$_{2}$ can also be produced by nuclear detonations.) The vast majority of burnable organic matter is directly or indirectly derived from oxygenic photosynthesis, meaning an abiotic or anoxic world would not have abundant preserved organic matter. To burn this biomass, one needs an atmosphere with oxygen. The observation of high abundances of NO$_{2}$ on an exoplanet atmosphere would indicate a sustained source of industrial production, likely requiring an oxic atmosphere and indicating a significant source of biomass to sustain long-term industrial activity. While NO$_{2}$ can exist in abundant quantities on planets around K-dwarf stars, it may not necessarily be a desirable thing for the inhabitants if they have biology similar to humans, because exposure to NO$_{2}$ could cause impairment of lung function and/or recurrent respiratory problems \citep{faustini2014nitrogen}. Conversely, if extraterrestrial biology is sufficiently different from Earth life, then it could be impervious to NO$_{2}$ toxicity. In this respect, NO$_{2}$ on K-dwarfs is similar to the likely accumulation of abiotic and biologically produced CO on Earth-like planets orbiting mid-to-late M-dwarfs, in addition to the accumulation of biosignature gases \citep{Eddie2019}. 
 
Missions like LUVOIR, HabEX, and OST may have biosignature targets as a priority, so it may be untenable to seek dedicated observing time for exclusive technosignature detection. However, in the search for exo-Earth candidates, we will undoubtedly detect other planets within the stellar system \citep{Stark2014, Kopparapu2018}. LUVOIR and HabEX will be able to simultaneously obtain the spectra of the other bright planets in the system, while performing their observations on a prime HZ target. Consequently, there may not be a need to schedule separate observation time for technosignature detection, as such efforts could ``piggy back'' on a routine survey to observe both HZ and non-HZ planets \citep{LL2019}. However, this assumes that the NO$_{2}$ detection will likely occur within the total integrated observational time spent on the prime HZ candidate for the biosignature detection, whereas Fig. \ref{fig:snr} indicates lower NO$_{2}$ abundances may require longer search times.

\section{Conclusion} \label{sec:conclusion}

The presence of NO$_{2}$ on Earth today results in part from sustained industrial processes in urban areas. This paper suggests that the detection of NO$_{2}$ in an exoplanet atmosphere could serve as a technosignature, as Earth-level biogenic sources would be unable to generate detectable atmospheric abundances of NO$_{2}$.  Using a 1-D photochemical model that uses present Earth atmospheric temperature profile, we find that it would be challenging to detect Earth-level NO$_{2}$ around  G and K-dwarf stars through direct imaging with only 10 hours of observation time. To detect present Earth-level NO$_{2}$ concentration with a SNR $\sim 5$, it would take $\sim 400$ hours of LUVOIR-15m telescope. Such large programs may be possible considering several hundred hours of observing time spent on Hubble UDF and CANDLES surveys. Historically, the United States NO$_{2}$ emission varied (gone down) by roughly a factor of $\sim 3$ over 40 years from 1980-2019. Hence, there might be a possibility to detect 40-year old Earth-level industrialized society with even less time. 
%The higher rate of NO$_{2}$ photolysis for Earth-like planets orbiting Sun-like stars means such targets may require 10 times the current NO$_{2}$ level in order to be detected with a 15 meter LUVOIR-A like telescope at a signal-to-noise ratio of 5 with 100 hours of observation time. 
In this cloud free model, habitable planets orbiting K-dwarf stars, by comparison, would marginally need less amount of time to detect present-day NO$_{2}$ abundance. The advantage of searching K-dwarf planets has already been noted in the search for biosignatures \citep{cuntz2016exobiology,Arney2019}, and our results indicate that K-dwarf planets could similarly be advantageous when searching for technosignatures like NO$_{2}$.

However, when we prescribe water-ice and liquid water clouds, there is a moderate decrease in the SNR of the geometric albedo spectrum from LUVOIR-15\,m, with present Earth-level NO$_{2}$ concentration on an Earth-like planet around a Sun-like star at 10\,pc. Clouds and aerosols can reduce the detectability and could mimic the NO$_{2}$ feature, posing a challenge to the unique identification of this signature. 
This highlights the need for performing these calculations with a 3-D climate model which can simulate variability of the cloud cover and atmospheric dynamics  self-consistently. 

While NO${_2}$ absorbs even in the near-IR and mid-IR, we find that transit observations in this region with JWST and OST may prove challenging to detect NO$_{2}$ because of the weaker absorption and also due to overlapping gas absorption of potent greenhouse gases such as H$_{2}$O, CO$_{2}$ and CH$_{4}$.

Further work is needed to explore the detectability of NO$_{2}$ on Earth-like planets around M-dwarfs in direct imaging observations in the near-IR  with ground-based 30\,m class telescopes. NO$_{2}$ concentrations increase on planets around cooler stars due to reduced availability of short-wavelength photons that can  photolyze  NO$_{2}$.
Non-detectability at longer observation times could place upper limits on the amount NO$_{2}$ present on M-dwarf HZ planets like  Prox Cen b. 

The serendipitous detection of NO$_{2}$, or any other potential artificial atmospheric spectral signature (CFCs, for example) may become a watershed event in the search for life (biological or technological). Is it likely that biosignatures are more prevalent than technosignatures? We will not know for certain until we search. Our aim in this study is to point out that both biosignatures and technosignatures are two sides of the same coin, and the search for both can co-exist together with upcoming observatories. It is worth pointing out the obvious in this concluding statement: the question ``Are we alone?''---which has been the driving force behind the search for extraterrestrial biosignatures---is a question posed by a technological civilization. 

%\begin{deluxetable*}{cchlDlc}
%\tablenum{1}
%\tablecaption{Fun facts about the first 10 messier objects\label{tab:messier}}
%\tablewidth{0pt}
%\tablehead{
%\colhead{Messier} & \colhead{NGC/IC} & \nocolhead{Common} & \colhead{Object} &
%\multicolumn2c{Distance} & \colhead{} & \colhead{V} \\
%\colhead{Number} & \colhead{Number} & \nocolhead{Name} & \colhead{Type} &
%\multicolumn2c{(kpc)} & \colhead{Constellation} & \colhead{(mag)}
%}
%\decimalcolnumbers
%\startdata
%M1 & NGC 1952 & Crab Nebula & Supernova remnant & 2 & Taurus & 8.4 \\
%M2 & NGC 7089 & Messier 2 & Cluster, globular & 11.5 & Aquarius & 6.3 \\
%M3 & NGC 5272 & Messier 3 & Cluster, globular & 10.4 & Canes Venatici &  6.2 \\
%M4 & NGC 6121 & Messier 4 & Cluster, globular & 2.2 & Scorpius & 5.9 \\
%M5 & NGC 5904 & Messier 5 & Cluster, globular & 24.5 & Serpens & 5.9 \\
%M6 & NGC 6405 & Butterfly Cluster & Cluster, open & 0.31 & Scorpius & 4.2 \\
%M7 & NGC 6475 & Ptolemy Cluster & Cluster, open & 0.3 & Scorpius & 3.3 \\
%M8 & NGC 6523 & Lagoon Nebula & Nebula with cluster & 1.25 & Sagittarius & 6.0 \\
%M9 & NGC 6333 & Messier 9 & Cluster, globular & 7.91 & Ophiuchus & 8.4 \\
%M10 & NGC 6254 & Messier 10 & Cluster, globular & 4.42 & Ophiuchus & 6.4 \\
%\enddata
%\tablecomments{This table ``hides'' the third column in the \latex\ when compiled.
%The Distance is also centered on the decimals.  Note that when using decimal
%alignment you need to include the {\tt\string\decimals} command before
%{\tt\string\startdata} and all of the values in that column have to have a
%space before the next ampersand.}
%\end{deluxetable*}

\acknowledgments

The authors would like to thank an anonymous reviewer whose comments greatly improved the manuscript. We would also like to thank Sandra Bastelberger, Chester ``Sonny'' Harman, Thomas Fauchez, James Kasting and Avi Mandell for discussions that helped in this work. R. K. would like to acknowledge Vivaswan Kopparapu, his 11 year old son, who helped R. K. to realize that increasing the LUVOIR-15\,m observation time by a factor of 4 doubles the NO$_{2}$ SNR for an Earth-Sun system at 10\,pc. Goddard affiliates acknowledge support from the GSFC Sellers Exoplanet Environments Collaboration (SEEC), which is supported by NASA’s Planetary Science Division’s Research Program.
J.H.M. gratefully acknowledges support from the NASA Exobiology program under grant 80NSSC20K0622.
This work was performed as part of NASA's Virtual Planetary Laboratory,
supported by the National Aeronautics and Space Administration through the
NASA Astrobiology Institute under solicitation NNH12ZDA002C and
Cooperative Agreement Number NNA13AA93A, and by the NASA Astrobiology
Program under grant 80NSSC18K0829 as part of the Nexus for Exoplanet
System Science (NExSS) research coordination network.

\appendix
\section{Comparison of LUVOIR Noise Models}
\label{appendix:cg}

% Intro to section
We conducted a comparison between the LUVOIR coronagraph noise model included in PSG and the Python implementation of the \citet{Robinson2016} coronagraph noise model from \citet{Lustig-Yaeger2019cg} (henceforth CG). We found the two models to agree very well (Figure \ref{fig:cg_compare}), with both models implementing very similar formalisms for computing sensitivities. 

\begin{figure*}[t!]
\centering
\includegraphics[width=0.95\textwidth]{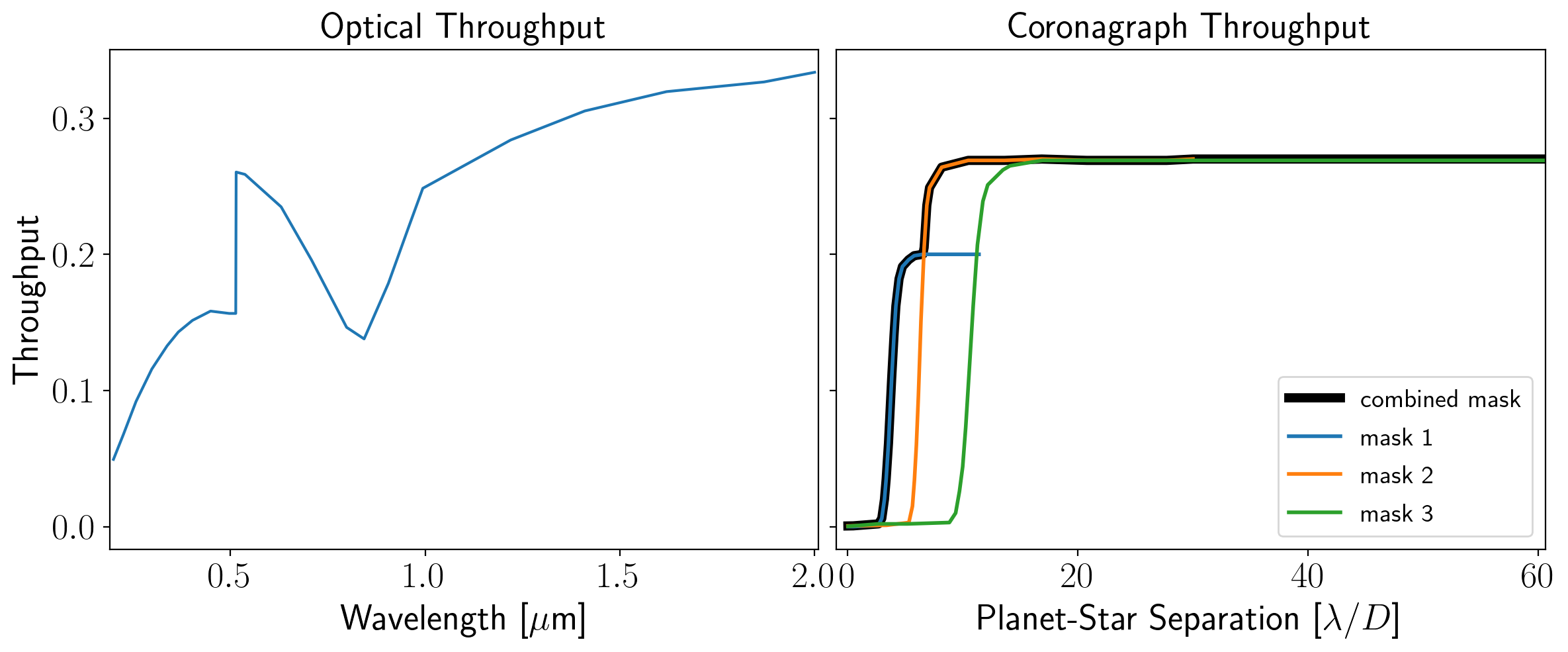}
\caption{Coronagraph throughputs used for LUVOIR-A noise modeling. The left panel shows the wavelength dependent optical throughput. The right panel shows the coronagraph throughput as a function of planet-star separation.}  
\label{fig:cg_tput}
\end{figure*}

\begin{figure*}[t!]
\centering
\includegraphics[width=0.95\textwidth]{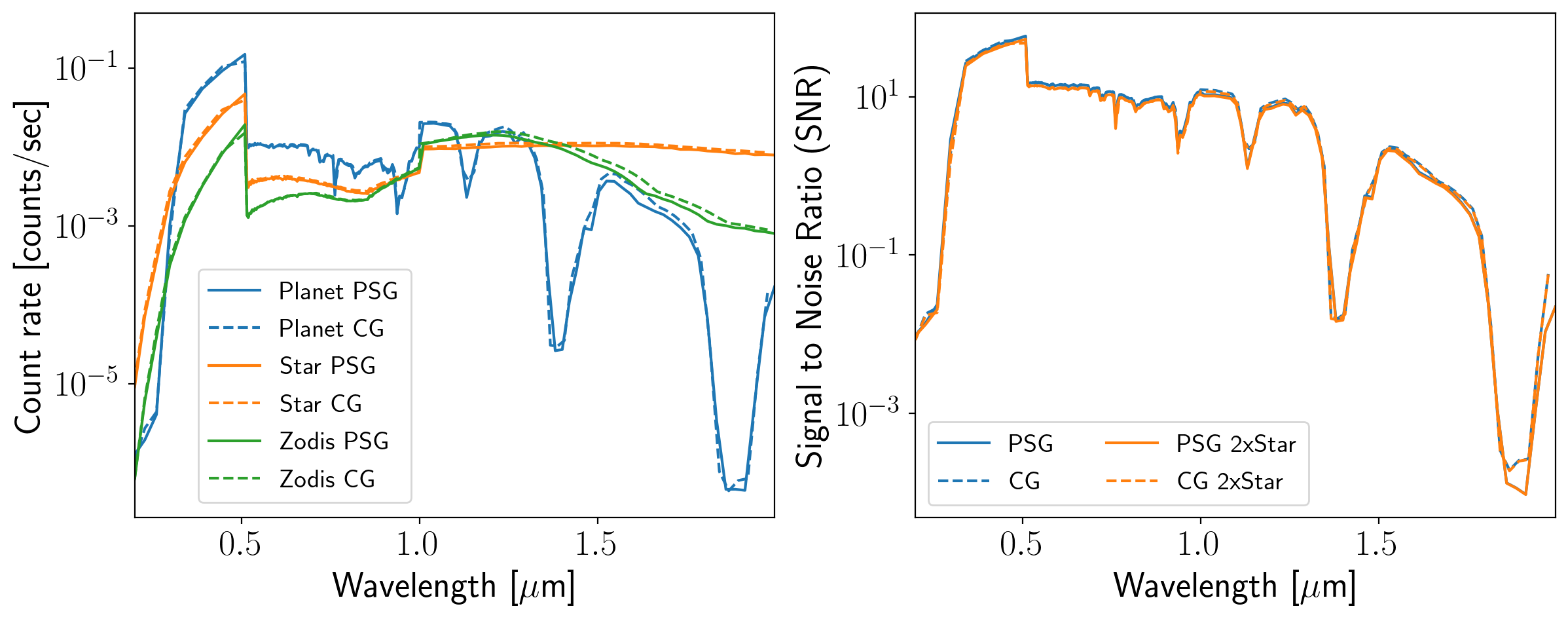}
\caption{Comparison between PSG and CG for LUVOIR-A. Left: photon count rates for the planet signal and dominant noise sources. Right: Precision on the visible spectrum in a 10 hour exposure for PSG and CG. The 2xStar values present the case in which star leakage is also considered in the ``off'' position. Both panels were computed assuming the native spectral resolutions of the channels (RP=7/140/70 for UV/VIS/NIR respectively).}  
\label{fig:cg_compare}
\end{figure*}

% Description of assumed throughputs 
We define the end-to-end throughput for the planetary fluxes as: $T_{total} = T_{Tele} \times T_{cor} \times T_{opt} \times T_{read} \times T_{QE}$, where $T_{Tele}$ accounts for light lost due to contamination and inefficiencies in the main collecting area, $T_{cor}$ is the coronagraphic throughput at this planet-star separation, $T_{opt}$ is the optical throughput (the transmissivity of all optics), $T_{QE}$ is the raw quantum efficiency (QE) of the detector, and $T_{read}$ is the read-out efficiencies. The left panel in Figure \ref{fig:cg_tput} shows the optical throughput ($T_{opt}$) from \citet{Stark2019} and the right panel shows the coronagraph throughput as a function of planet-star separation ($T_{cor}$). Although the design of the LUVOIR-A coronagraph has multiple different masks with slightly different IWAs, both coronagraph models use a combined mask (shown in Figure \ref{fig:cg_tput}) to approximate the optimal use of the coronagraph for any simulated target. Importantly, the coronagraph throughput already accounts for the fraction of the exoplanetary light that falls within the photometric aperture, denoted $f_{pa}$ in \citet{Robinson2016}, so we manually set $f_{pa}=1$ in the CG model to properly account for this factor. The number of stellar photons is defined by the contrast at the core throughput, and thus the number  of stellar photons is calculated as $C \cdot \mathrm{max}(T_{cor}) \approx 10^{-10} \cdot 0.27$.    
For $T_{Tele}$, we adopt 0.95 for all wavelengths, on par with the particulate coverage fraction for JWST’s mirrors. EMCCD detectors are expected to have $T_{read}$ near 0.75 \citep{Stark2019}, while for NIR and other detectors, read-out inefficiencies and bad-pixels may account to a similar value, and we adopt $T_{read}$=0.75 across all detectors as a conservative estimate. The reported quantum efficiency of the different detectors ranges from 0.6 to 0.9, yet technological improvements in several of these detectors could be expected in the near future, and we adopt a general $T_{QE}$=0.9 for all detectors.

The Signal-to-Noise ratio (SNR) is effectively defined by the different sources of noise, quantified as count rates. We not only compared resulting SNRs between the two models, but also the simulated count rates for the different components, and found very good agreement (Figure \ref{fig:cg_compare}). For these simulations, we assumed a circular aperture defined by diffraction (1.22 $\lambda/D$), an exo-zodiacal level of 4.5 times the one of our solar system (22 $mag/arcsec^{2}$), and a local zodi level of 22.5 $mag/arcsec^{2}$. The noise term was computed as
\begin{equation}
\label{eqn:cg_noise}
    C_{noise} = \sqrt{C_p + C_s + 2C_b}
\end{equation}
where $C_p$ is the total number of planet photons, $C_s$ is the stellar photon noise (e.g. ``leakage'' through the coronagraph), and $C_b$ is the total background, which includes all other noise sources such as zodi, exozodi, dark current, thermal, and read noise. Observations are normally performed as on-off, meaning one with the planet, and one without. As such, the background sources of noise need to be counted twice (equation \ref{eqn:cg_noise}). Depending on the observational procedure, the stellar photons can be assumed to be present in the ``off'' position or not. \citet{Robinson2016} assumes by default that the star is also in the ``off'' position, and therefore doubles $C_s$, while the default in PSG is the ``off'' position without star leakage (so only counted once, equation \ref{eqn:cg_noise}). We explored the impact on the SNR of this assumption in the observational procedure, and only observe small ($<$10\%) differences in the resulting SNR (Figure \ref{fig:cg_compare}). 

% Throughput plot(s) 

% Coronagraph comparison

\begin{deluxetable}{llc}
%\tablefontsize{\scriptsize}
\tablewidth{0.95\linewidth}
%\tablewidth{\columnwidth}
\tablecaption{LUVOIR-A coronagraph model input parameters}
\tablehead{\colhead{Parameter} & \colhead{Description} & \colhead{Value}}
\startdata
$D$ & Mirror Diameter & 15 m \\ 
$C$ & Contrast & $10^{-10}$ \\
$T_{opt}$ & Optical Throughput & Figure \ref{fig:cg_tput} (left) \\ 
$T_{cor}$ & Coronagraph Throughput & Figure \ref{fig:cg_tput} (right) \\
$R_{e-}$ & Read Noise (UVIS/NIR) & 0 / 2.5 \\
$D_{e-}$ & Dark Current (UVIS/NIR) & 3E-5 / 2E-3 $s^{-1}$ \\ 
$X$ & Circular Photometric Aperture Radius & $0.61 \lambda / D$ \\  
$N_{ez}$ & Number of Exozodis & 4.5
\enddata
\end{deluxetable}

%% To help institutions obtain information on the effectiveness of their 
%% telescopes the AAS Journals has created a group of keywords for telescope 
%% facilities.
%
%% Following the acknowledgments section, use the following syntax and the
%% \facility{} or \facilities{} macros to list the keywords of facilities used 
%% in the research for the paper.  Each keyword is check against the master 
%% list during copy editing.  Individual instruments can be provided in 
%% parentheses, after the keyword, but they are not verified.

\vspace{5mm}
%\facilities{HST(STIS), Swift(XRT and UVOT), AAVSO, CTIO:1.3m,
%CTIO:1.5m,CXO}

%% Similar to \facility{}, there is the optional \software command to allow 
%% authors a place to specify which programs were used during the creation of 
%% the manuscript. Authors should list each code and include either a
%% citation or url to the code inside ()s when available.

%\software{astropy \citep{2013A&A...558A..33A},  
%          Cloudy \citep{2013RMxAA..49..137F}, 
 %         SExtractor \citep{1996A&AS..117..393B}
 %         }

%% Appendix material should be preceded with a single \appendix command.
%% There should be a \section command for each appendix. Mark appendix
%% subsections with the same markup you use in the main body of the paper.

%% Each Appendix (indicated with \section) will be lettered A, B, C, etc.
%% The equation counter will reset when it encounters the \appendix
%% command and will number appendix equations (A1), (A2), etc. The
%% Figure and Table counter will not reset.

%\appendix

%% For this sample we use BibTeX plus aasjournals.bst to generate the
%% the bibliography. The sample63.bib file was populated from ADS. To
%% get the citations to show in the compiled file do the following:
%%
%% pdflatex sample63.tex
%% bibtext sample63
%% pdflatex sample63.tex
%% pdflatex sample63.tex

\bibliography{sample63}{}

\begin{thebibliography}{}
\expandafter\ifx\csname natexlab\endcsname\relax\def\natexlab#1{#1}\fi
\providecommand{\url}[1]{\href{#1}{#1}}
\providecommand{\dodoi}[1]{doi:~\href{http://doi.org/#1}{\nolinkurl{#1}}}
\providecommand{\doeprint}[1]{\href{http://ascl.net/#1}{\nolinkurl{http://ascl.net/#1}}}
\providecommand{\doarXiv}[1]{\href{https://arxiv.org/abs/#1}{\nolinkurl{https://arxiv.org/abs/#1}}}

\bibitem[{{Arney} {et~al.}(2016){Arney}, {Domagal-Goldman}, {Meadows}, {Wolf},
  {Schwieterman}, {Charnay}, {Claire}, {H{\'e}brard}, \& {Trainer}}]{Arney2016}
{Arney}, G., {Domagal-Goldman}, S.~D., {Meadows}, V.~S., {et~al.} 2016,
  Astrobiology, 16, 873, \dodoi{10.1089/ast.2015.1422}

\bibitem[{{Arney}(2019)}]{Arney2019}
{Arney}, G.~N. 2019, \apjl, 873, L7, \dodoi{10.3847/2041-8213/ab0651}

\bibitem[{Arnold(2005)}]{Arnold05}
Arnold, L. 2005, $\backslash$apj, 627, 534, \dodoi{10.1086/430437}

\bibitem[{Bauwens {et~al.}(2020)Bauwens, Compernolle, Stavrakou, M{\"u}ller,
  van Gent, Eskes, Levelt, van~der A, Veefkind, Vlietinck,
  {et~al.}}]{bauwensimpact}
Bauwens, M., Compernolle, S., Stavrakou, T., {et~al.} 2020, Geophysical
  Research Letters, e2020GL087978

\bibitem[{Beckwith {et~al.}(2006)Beckwith, Stiavelli, Koekemoer, Caldwell,
  Ferguson, Hook, Lucas, Bergeron, Corbin, Jogee,
  {et~al.}}]{beckwith2006hubble}
Beckwith, S.~V., Stiavelli, M., Koekemoer, A.~M., {et~al.} 2006, The
  Astronomical Journal, 132, 1729

\bibitem[{{Benneke} {et~al.}(2019){Benneke}, {Wong}, {Piaulet}, {Knutson},
  {Lothringer}, {Morley}, {Crossfield}, {Gao}, {Greene}, {Dressing},
  {Dragomir}, {Howard}, {McCullough}, {Kempton}, {Fortney}, \&
  {Fraine}}]{paper22019}
{Benneke}, B., {Wong}, I., {Piaulet}, C., {et~al.} 2019, \apjl, 887, L14,
  \dodoi{10.3847/2041-8213/ab59dc}

\bibitem[{Bracewell(1960)}]{BRACEWELL1960}
Bracewell, R.~N. 1960, Nature, 186, 670, \dodoi{10.1038/186670a0}

\bibitem[{{Carrigan Jr.}(2009)}]{carrigan09b}
{Carrigan Jr.}, R. 2009, in Astronomical Society of the Pacific Conference
  Series, Vol. 420, Bioastronomy 2007: Molecules, Microbes and Extraterrestrial
  Life, ed. K.~Meech, J.~Keane, M.~Mumma, J.~Siefert, \& D.~Werthimer, 415

\bibitem[{{Catling} {et~al.}(2018){Catling}, {Krissansen-Totton}, {Kiang},
  {Crisp}, {Robinson}, {DasSarma}, {Rushby}, {Del Genio}, {Bains}, \&
  {Domagal-Goldman}}]{Catling2018}
{Catling}, D.~C., {Krissansen-Totton}, J., {Kiang}, N.~Y., {et~al.} 2018,
  Astrobiology, 18, 709, \dodoi{10.1089/ast.2017.1737}

\bibitem[{{Chance} \& {Kurucz}(2010)}]{CK2010}
{Chance}, K., \& {Kurucz}, R.~L. 2010, \jqsrt, 111, 1289,
  \dodoi{10.1016/j.jqsrt.2010.01.036}

\bibitem[{{Charbonneau} {et~al.}(2002){Charbonneau}, {Brown}, {Noyes}, \&
  {Gilliland}}]{dave2002}
{Charbonneau}, D., {Brown}, T.~M., {Noyes}, R.~W., \& {Gilliland}, R.~L. 2002,
  \apj, 568, 377, \dodoi{10.1086/338770}

\bibitem[{Crutzen(2006)}]{crutzen2006anthropocene}
Crutzen, P.~J. 2006, in Earth system science in the anthropocene (Springer),
  13--18

\bibitem[{Cuntz \& Guinan(2016)}]{cuntz2016exobiology}
Cuntz, M., \& Guinan, E. 2016, The Astrophysical Journal, 827, 79

\bibitem[{{Defr{\`e}re} {et~al.}(2018){Defr{\`e}re}, {L{\'e}ger}, {Absil},
  {Beichman}, {Biller}, {Danchi}, {Ergenzinger}, {Eiroa}, {Ertel}, {Fridlund},
  {Mu{\~n}oz}, {Gillon}, {Glasse}, {Godolt}, {Grenfell}, {Kraus}, {Labadie},
  {Lacour}, {Liseau}, {Martin}, {Mennesson}, {Micela}, {Minardi}, {Quanz},
  {Rauer}, {Rinehart}, {Santos}, {Selsis}, {Surdej}, {Tian}, {Villaver},
  {Wheatley}, \& {Wyatt}}]{life2018}
{Defr{\`e}re}, D., {L{\'e}ger}, A., {Absil}, O., {et~al.} 2018, Experimental
  Astronomy, 46, 543, \dodoi{10.1007/s10686-018-9613-2}

\bibitem[{{Domagal-Goldman} {et~al.}(2014){Domagal-Goldman}, {Segura},
  {Claire}, {Robinson}, \& {Meadows}}]{Domagal-Goldman2014}
{Domagal-Goldman}, S.~D., {Segura}, A., {Claire}, M.~W., {Robinson}, T.~D., \&
  {Meadows}, V.~S. 2014, \apj, 792, 90, \dodoi{10.1088/0004-637X/792/2/90}

\bibitem[{{Dyson}(1960)}]{dyson60}
{Dyson}, F.~J. 1960, Science, 131, 1667, \dodoi{10.1126/science.131.3414.1667}

\bibitem[{Faustini {et~al.}(2014{\natexlab{a}})Faustini, Rapp, \&
  Forastiere}]{Faustini2014}
Faustini, A., Rapp, R., \& Forastiere, F. 2014{\natexlab{a}}, European
  Respiratory Journal, 44, 744

\bibitem[{Faustini {et~al.}(2014{\natexlab{b}})Faustini, Rapp, \&
  Forastiere}]{faustini2014nitrogen}
---. 2014{\natexlab{b}}, European Respiratory Journal, 44, 744

\bibitem[{Forgan(2013)}]{Forgan13}
Forgan, D. 2013, Journal of the British Interplanetary Society, 66, 144

\bibitem[{{France} {et~al.}(2016){France}, {Loyd}, {Youngblood}, {Brown},
  {Schneider}, {Hawley}, {Froning}, {Linsky}, {Roberge}, {Buccino},
  {Davenport}, {Fontenla}, {Kaltenegger}, {Kowalski}, {Mauas}, {Miguel},
  {Redfield}, {Rugheimer}, {Tian}, {Vieytes}, {Walkowicz}, \&
  {Weisenburger}}]{France2016}
{France}, K., {Loyd}, R.~O.~P., {Youngblood}, A., {et~al.} 2016, \apj, 820, 89,
  \dodoi{10.3847/0004-637X/820/2/89}

\bibitem[{Frank {et~al.}(2017)Frank, Kleidon, \& Alberti}]{frank2017earth}
Frank, A., Kleidon, A., \& Alberti, M. 2017, Anthropocene, 19, 13

\bibitem[{Freitas~Jr \& Valdes(1980)}]{freitas1980search}
Freitas~Jr, R.~A., \& Valdes, F. 1980, Icarus, 42, 442

\bibitem[{{Fujii} {et~al.}(2018){Fujii}, {Angerhausen}, {Deitrick},
  {Domagal-Goldman}, {Grenfell}, {Hori}, {Kane}, {Pall{\'e}}, {Rauer},
  {Siegler}, {Stapelfeldt}, \& {Stevenson}}]{Fujii2018}
{Fujii}, Y., {Angerhausen}, D., {Deitrick}, R., {et~al.} 2018, Astrobiology,
  18, 739, \dodoi{10.1089/ast.2017.1733}

\bibitem[{Grenfell(2017)}]{grenfell2017review}
Grenfell, J.~L. 2017, Physics Reports, 713, 1

\bibitem[{{Grogin} {et~al.}(2011){Grogin}, {Kocevski}, {Faber}, {Ferguson},
  {Koekemoer}, {Riess}, {Acquaviva}, {Alexander}, {Almaini}, {Ashby}, {Barden},
  {Bell}, {Bournaud}, {Brown}, {Caputi}, {Casertano}, {Cassata}, {Castellano},
  {Challis}, {Chary}, {Cheung}, {Cirasuolo}, {Conselice}, {Roshan Cooray},
  {Croton}, {Daddi}, {Dahlen}, {Dav{\'e}}, {de Mello}, {Dekel}, {Dickinson},
  {Dolch}, {Donley}, {Dunlop}, {Dutton}, {Elbaz}, {Fazio}, {Filippenko},
  {Finkelstein}, {Fontana}, {Gardner}, {Garnavich}, {Gawiser}, {Giavalisco},
  {Grazian}, {Guo}, {Hathi}, {H{\"a}ussler}, {Hopkins}, {Huang}, {Huang},
  {Jha}, {Kartaltepe}, {Kirshner}, {Koo}, {Lai}, {Lee}, {Li}, {Lotz}, {Lucas},
  {Madau}, {McCarthy}, {McGrath}, {McIntosh}, {McLure}, {Mobasher},
  {Moustakas}, {Mozena}, {Nandra}, {Newman}, {Niemi}, {Noeske}, {Papovich},
  {Pentericci}, {Pope}, {Primack}, {Rajan}, {Ravindranath}, {Reddy}, {Renzini},
  {Rix}, {Robaina}, {Rodney}, {Rosario}, {Rosati}, {Salimbeni}, {Scarlata},
  {Siana}, {Simard}, {Smidt}, {Somerville}, {Spinrad}, {Straughn}, {Strolger},
  {Telford}, {Teplitz}, {Trump}, {van der Wel}, {Villforth}, {Wechsler},
  {Weiner}, {Wiklind}, {Wild}, {Wilson}, {Wuyts}, {Yan}, \& {Yun}}]{grogin2011}
{Grogin}, N.~A., {Kocevski}, D.~D., {Faber}, S.~M., {et~al.} 2011, \apjs, 197,
  35, \dodoi{10.1088/0067-0049/197/2/35}

\bibitem[{{Haqq-Misra} \& {Kopparapu}(2012)}]{JR2012}
{Haqq-Misra}, J., \& {Kopparapu}, R.~K. 2012, Acta Astronautica, 72, 15,
  \dodoi{10.1016/j.actaastro.2011.10.010}

\bibitem[{{Haqq-Misra} {et~al.}(2020){Haqq-Misra}, {Kopparapu}, \&
  {Schwieterman}}]{jacob2020}
{Haqq-Misra}, J., {Kopparapu}, R.~K., \& {Schwieterman}, E. 2020, Astrobiology,
  20, 572, \dodoi{10.1089/ast.2019.2154}

\bibitem[{{Harman} {et~al.}(2018){Harman}, {Felton}, {Hu}, {Domagal-Goldman},
  {Segura}, {Tian}, \& {Kasting}}]{Harman2018}
{Harman}, C.~E., {Felton}, R., {Hu}, R., {et~al.} 2018, \apj, 866, 56,
  \dodoi{10.3847/1538-4357/aadd9b}

\bibitem[{{Harman} {et~al.}(2015){Harman}, {Schwieterman}, {Schottelkotte}, \&
  {Kasting}}]{Harman2015}
{Harman}, C.~E., {Schwieterman}, E.~W., {Schottelkotte}, J.~C., \& {Kasting},
  J.~F. 2015, \apj, 812, 137, \dodoi{10.1088/0004-637X/812/2/137}

\bibitem[{Holmes {et~al.}(2013)Holmes, Prather, S{\o}vde, \&
  Myhre}]{Holmes2013}
Holmes, C.~D., Prather, M.~J., S{\o}vde, O.~A., \& Myhre, G. 2013, Atmospheric
  Chemistry and Physics, 13, 285, \dodoi{10.5194/acp-13-285-2013}

\bibitem[{Kaltenegger(2017)}]{kaltenegger2017characterize}
Kaltenegger, L. 2017, Annual Review of Astronomy and Astrophysics, 55, 433

\bibitem[{Kasting \& Ackerman(1985)}]{KA1985}
Kasting, J., \& Ackerman, T. 1985, Journal of atmospheric chemistry, 3, 321

\bibitem[{{Kasting} {et~al.}(1979){Kasting}, {Liu}, \& {Donahue}}]{Kasting79}
{Kasting}, J.~F., {Liu}, S.~C., \& {Donahue}, T.~M. 1979, \jgr, 84, 3097,
  \dodoi{10.1029/JC084iC06p03097}

\bibitem[{Kipping \& Teachey(2016)}]{Kipping16}
Kipping, D., \& Teachey, A. 2016, $\backslash$mnras, 459, 1233,
  \dodoi{10.1093/mnras/stw672}

\bibitem[{{Kopparapu} {et~al.}(2012){Kopparapu}, {Kasting}, \&
  {Zahnle}}]{Kopparapu2012}
{Kopparapu}, R.~k., {Kasting}, J.~F., \& {Zahnle}, K.~J. 2012, \apj, 745, 77,
  \dodoi{10.1088/0004-637X/745/1/77}

\bibitem[{{Kopparapu} {et~al.}(2014){Kopparapu}, {Ramirez}, {SchottelKotte},
  {Kasting}, {Domagal-Goldman}, \& {Eymet}}]{Kopparapu2014}
{Kopparapu}, R.~K., {Ramirez}, R.~M., {SchottelKotte}, J., {et~al.} 2014,
  \apjl, 787, L29, \dodoi{10.1088/2041-8205/787/2/L29}

\bibitem[{{Kopparapu} {et~al.}(2013){Kopparapu}, {Ramirez}, {Kasting}, {Eymet},
  {Robinson}, {Mahadevan}, {Terrien}, {Domagal-Goldman}, {Meadows}, \&
  {Deshpande}}]{Kopparapu2013}
{Kopparapu}, R.~K., {Ramirez}, R., {Kasting}, J.~F., {et~al.} 2013, \apj, 765,
  131, \dodoi{10.1088/0004-637X/765/2/131}

\bibitem[{{Kopparapu} {et~al.}(2018){Kopparapu}, {H{\'e}brard}, {Belikov},
  {Batalha}, {Mulders}, {Stark}, {Teal}, {Domagal-Goldman}, \&
  {Mandell}}]{Kopparapu2018}
{Kopparapu}, R.~K., {H{\'e}brard}, E., {Belikov}, R., {et~al.} 2018, \apj, 856,
  122, \dodoi{10.3847/1538-4357/aab205}

\bibitem[{Kraus \& Hofzumahaus(1998)}]{kraus1998}
Kraus, A., \& Hofzumahaus, A. 1998, in Atmospheric Measurements during
  POPCORN—Characterisation of the Photochemistry over a Rural Area
  (Springer), 161--180

\bibitem[{{Krissansen-Totton} {et~al.}(2018){Krissansen-Totton}, {Olson}, \&
  {Catling}}]{Josh2018}
{Krissansen-Totton}, J., {Olson}, S., \& {Catling}, D.~C. 2018, Science
  Advances, 4, eaao5747, \dodoi{10.1126/sciadv.aao5747}

\bibitem[{Kuhn \& Berdyugina(2015)}]{kuhn2015global}
Kuhn, J.~R., \& Berdyugina, S.~V. 2015, International journal of astrobiology,
  14, 401

\bibitem[{Lammer {et~al.}(2019)Lammer, Spro{\ss}, Grenfell, Scherf, Fossati,
  Lendl, \& Cubillos}]{lammer2019role}
Lammer, H., Spro{\ss}, L., Grenfell, J.~L., {et~al.} 2019, Astrobiology, 19,
  927

\bibitem[{Lamsal {et~al.}(2013)Lamsal, Martin, Parrish, \&
  Krotkov}]{lamsal2013scaling}
Lamsal, L., Martin, R., Parrish, D., \& Krotkov, N. 2013, Environmental science
  \& technology, 47, 7855

\bibitem[{Lewis \& Maslin(2015)}]{lewis2015defining}
Lewis, S.~L., \& Maslin, M.~A. 2015, Nature, 519, 171

\bibitem[{{Lin} {et~al.}(2014){Lin}, {Gonzalez Abad}, \& {Loeb}}]{Lin2014}
{Lin}, H.~W., {Gonzalez Abad}, G., \& {Loeb}, A. 2014, \apjl, 792, L7,
  \dodoi{10.1088/2041-8205/792/1/L7}

\bibitem[{{Lincowski} {et~al.}(2018){Lincowski}, {Meadows}, {Crisp},
  {Robinson}, {Luger}, {Lustig-Yaeger}, \& {Arney}}]{Lincowski2018}
{Lincowski}, A.~P., {Meadows}, V.~S., {Crisp}, D., {et~al.} 2018, \apj, 867,
  76, \dodoi{10.3847/1538-4357/aae36a}

\bibitem[{Lingam \& Loeb(2017)}]{Lingam17}
Lingam, M., \& Loeb, A. 2017, $\backslash$mnras, 470, L82,
  \dodoi{10.1093/mnrasl/slx084}

\bibitem[{{Lingam} \& {Loeb}(2019)}]{LL2019}
{Lingam}, M., \& {Loeb}, A. 2019, Astrobiology, 19, 28,
  \dodoi{10.1089/ast.2018.1936}

\bibitem[{Loeb \& Turner(2012)}]{Loeb11}
Loeb, A., \& Turner, E. 2012, Astrobiology, 12, 290,
  \dodoi{10.1089/ast.2011.0758}

\bibitem[{{Loyd} {et~al.}(2016){Loyd}, {France}, {Youngblood}, {Schneider},
  {Brown}, {Hu}, {Linsky}, {Froning}, {Redfield}, {Rugheimer}, \&
  {Tian}}]{Loyd2016}
{Loyd}, R.~O.~P., {France}, K., {Youngblood}, A., {et~al.} 2016, \apj, 824,
  102, \dodoi{10.3847/0004-637X/824/2/102}

\bibitem[{{Lustig-Yaeger} {et~al.}(2019){Lustig-Yaeger}, {Meadows}, \&
  {Lincowski}}]{Jake2019}
{Lustig-Yaeger}, J., {Meadows}, V.~S., \& {Lincowski}, A.~P. 2019, \aj, 158,
  27, \dodoi{10.3847/1538-3881/ab21e0}

\bibitem[{Lustig-Yaeger {et~al.}(2019)Lustig-Yaeger, Robinson, \&
  Arney}]{Lustig-Yaeger2019cg}
Lustig-Yaeger, J., Robinson, T., \& Arney, G. 2019, Journal of Open Source
  Software, 4, 1387, \dodoi{10.21105/joss.01387}

\bibitem[{{Meadows} {et~al.}(2018){Meadows}, {Reinhard}, {Arney}, {Parenteau},
  {Schwieterman}, {Domagal-Goldman}, {Lincowski}, {Stapelfeldt}, {Rauer},
  {DasSarma}, {Hegde}, {Narita}, {Deitrick}, {Lustig-Yaeger}, {Lyons},
  {Siegler}, \& {Grenfell}}]{Meadows2018}
{Meadows}, V.~S., {Reinhard}, C.~T., {Arney}, G.~N., {et~al.} 2018,
  Astrobiology, 18, 630, \dodoi{10.1089/ast.2017.1727}

\bibitem[{Meadows {et~al.}(2018)Meadows, Arney, Schwieterman, Lustig-Yaeger,
  Lincowski, Robinson, Domagal-Goldman, Deitrick, Barnes, Fleming,
  {et~al.}}]{Meadows2018b}
Meadows, V.~S., Arney, G.~N., Schwieterman, E.~W., {et~al.} 2018, Astrobiology,
  18, 133

\bibitem[{O’Malley-James \& Kaltenegger(2019)}]{o2019expanding}
O’Malley-James, J.~T., \& Kaltenegger, L. 2019, The Astrophysical Journal
  Letters, 879, L20

\bibitem[{{Pall{\'e}}(2018)}]{palle2018}
{Pall{\'e}}, E. 2018, {The Detectability of Earth's Biosignatures Across Time},
  70, \dodoi{10.1007/978-3-319-55333-7_70}

\bibitem[{{Quanz} {et~al.}(2018){Quanz}, {Kammerer}, {Defr{\`e}re}, {Absil},
  {Glauser}, \& {Kitzmann}}]{quanz2018}
{Quanz}, S.~P., {Kammerer}, J., {Defr{\`e}re}, D., {et~al.} 2018, in Society of
  Photo-Optical Instrumentation Engineers (SPIE) Conference Series, Vol. 10701,
  Optical and Infrared Interferometry and Imaging VI, 107011I,
  \dodoi{10.1117/12.2312051}

\bibitem[{{Ranjan} {et~al.}(2020){Ranjan}, {Schwieterman}, {Harman}, {Fateev},
  {Sousa-Silva}, {Seager}, \& {Hu}}]{Ranjan2020}
{Ranjan}, S., {Schwieterman}, E.~W., {Harman}, C., {et~al.} 2020, \apj, 896,
  148, \dodoi{10.3847/1538-4357/ab9363}

\bibitem[{{Robinson} {et~al.}(2016){Robinson}, {Stapelfeldt}, \&
  {Marley}}]{Robinson2016}
{Robinson}, T.~D., {Stapelfeldt}, K.~R., \& {Marley}, M.~S. 2016, \pasp, 128,
  025003, \dodoi{10.1088/1538-3873/128/960/025003}

\bibitem[{Rose \& Wright(2004)}]{Rose2004}
Rose, C., \& Wright, G. 2004, Nature, 431, 47, \dodoi{10.1038/nature02884}

\bibitem[{Schneider(2010)}]{Schneider2010}
Schneider, J. 2010, Astrobiology, 10, 857, \dodoi{10.1089/ast.2010.9499}

\bibitem[{{Schwieterman} {et~al.}(2019){Schwieterman}, {Reinhard}, {Olson},
  {Harman}, \& {Lyons}}]{Eddie2019}
{Schwieterman}, E.~W., {Reinhard}, C.~T., {Olson}, S.~L., {Harman}, C.~E., \&
  {Lyons}, T.~W. 2019, \apj, 878, 19, \dodoi{10.3847/1538-4357/ab1d52}

\bibitem[{{Schwieterman} {et~al.}(2018){Schwieterman}, {Kiang}, {Parenteau},
  {Harman}, {DasSarma}, {Fisher}, {Arney}, {Hartnett}, {Reinhard}, {Olson},
  {Meadows}, {Cockell}, {Walker}, {Grenfell}, {Hegde}, {Rugheimer}, {Hu}, \&
  {Lyons}}]{Schwieterman2018}
{Schwieterman}, E.~W., {Kiang}, N.~Y., {Parenteau}, M.~N., {et~al.} 2018,
  Astrobiology, 18, 663, \dodoi{10.1089/ast.2017.1729}

\bibitem[{Seager {et~al.}(2012)Seager, Schrenk, \&
  Bains}]{seager2012astrophysical}
Seager, S., Schrenk, M., \& Bains, W. 2012, Astrobiology, 12, 61

\bibitem[{{Segura} {et~al.}(2005){Segura}, {Kasting}, {Meadows}, {Cohen},
  {Scalo}, {Crisp}, {Butler}, \& {Tinetti}}]{Segura2005}
{Segura}, A., {Kasting}, J.~F., {Meadows}, V., {et~al.} 2005, Astrobiology, 5,
  706, \dodoi{10.1089/ast.2005.5.706}

\bibitem[{Shklovskii \& Sagan(1966)}]{shklovskii1966}
Shklovskii, I.~S., \& Sagan, C. 1966, {Intelligent life in the universe:
  Vselennaja zizn'razum}, Delta-books (Holden-Day).
\newblock \url{https://books.google.com/books?id=o4cRAQAAIAAJ}

\bibitem[{{Stark} {et~al.}(2014){Stark}, {Roberge}, {Mandell}, \&
  {Robinson}}]{Stark2014}
{Stark}, C.~C., {Roberge}, A., {Mandell}, A., \& {Robinson}, T.~D. 2014, \apj,
  795, 122, \dodoi{10.1088/0004-637X/795/2/122}

\bibitem[{Stark {et~al.}(2019)Stark, Belikov, Bolcar, Cady, Crill, Ertel,
  Groff, Hildebrandt, Krist, Lisman, Mazoyer, Mennesson, Nemati, Pueyo,
  Rauscher, Riggs, Ruane, Shaklan, Sirbu, Soummer, Laurent, \&
  Zimmerman}]{Stark2019}
Stark, C.~C., Belikov, R., Bolcar, M.~R., {et~al.} 2019, Journal of
  Astronomical Telescopes, Instruments, and Systems, 5, 1 ,
  \dodoi{10.1117/1.JATIS.5.2.024009}

\bibitem[{Stevens {et~al.}(2016)Stevens, Forgan, \& James}]{Stevens16}
Stevens, A., Forgan, D., \& James, J. 2016, International Journal of
  Astrobiology, 15, 333, \dodoi{10.1017/S1473550415000397}

\bibitem[{{Suissa} {et~al.}(2020){Suissa}, {Wolf}, {Kopparapu}, {Villanueva},
  {Fauchez}, {Mandell}, {Arney}, {Gilbert}, {Schlieder}, {Barclay}, {Quintana},
  {Lopez}, {Rodriguez}, \& {Vanderburg}}]{gabby2020}
{Suissa}, G., {Wolf}, E.~T., {Kopparapu}, R.~k., {et~al.} 2020, arXiv e-prints,
  arXiv:2001.00955.
\newblock \doarXiv{2001.00955}

\bibitem[{{Technosignatures Workshop Participants}(2018)}]{tech2018}
{Technosignatures Workshop Participants}, N. 2018, arXiv e-prints,
  arXiv:1812.08681.
\newblock \doarXiv{1812.08681}

\bibitem[{{Tsiaras} {et~al.}(2019){Tsiaras}, {Waldmann}, {Tinetti}, {Tennyson},
  \& {Yurchenko}}]{paper12019}
{Tsiaras}, A., {Waldmann}, I.~P., {Tinetti}, G., {Tennyson}, J., \&
  {Yurchenko}, S.~N. 2019, Nature Astronomy, 3, 1086,
  \dodoi{10.1038/s41550-019-0878-9}

\bibitem[{{Villanueva} {et~al.}(2018){Villanueva}, {Smith}, {Protopapa},
  {Faggi}, \& {Mandell}}]{psg:2018}
{Villanueva}, G.~L., {Smith}, M.~D., {Protopapa}, S., {Faggi}, S., \&
  {Mandell}, A.~M. 2018, \jqsrt, 217, 86, \dodoi{10.1016/j.jqsrt.2018.05.023}

\bibitem[{{Walker} {et~al.}(2018){Walker}, {Bains}, {Cronin}, {DasSarma},
  {Danielache}, {Domagal-Goldman}, {Kacar}, {Kiang}, {Lenardic}, {Reinhard},
  {Moore}, {Schwieterman}, {Shkolnik}, \& {Smith}}]{Walker2018}
{Walker}, S.~I., {Bains}, W., {Cronin}, L., {et~al.} 2018, Astrobiology, 18,
  779, \dodoi{10.1089/ast.2017.1738}

\bibitem[{Whitmire \& Wright(1980)}]{Whitmire80}
Whitmire, D., \& Wright, D. 1980, $\backslash$icarus, 42, 149,
  \dodoi{10.1016/0019-1035(80)90253-5}

\bibitem[{{Wolf} {et~al.}(2019){Wolf}, {Kopparapu}, \& {Haqq-Misra}}]{wolf2019}
{Wolf}, E.~T., {Kopparapu}, R.~K., \& {Haqq-Misra}, J. 2019, \apj, 877, 35,
  \dodoi{10.3847/1538-4357/ab184a}

\bibitem[{{Wright}(2019)}]{wright2019}
{Wright}, J.~T. 2019, arXiv e-prints, arXiv:1907.07832.
\newblock \doarXiv{1907.07832}

\bibitem[{{Wright} {et~al.}(2016){Wright}, {Cartier}, {Zhao}, {Jontof-Hutter},
  \& {Ford}}]{GHAT4}
{Wright}, J.~T., {Cartier}, K.~M.~S., {Zhao}, M., {Jontof-Hutter}, D., \&
  {Ford}, E.~B. 2016, \apj, 816, 17, \dodoi{10.3847/0004-637X/816/1/17}

\bibitem[{{Wright} {et~al.}(2014){Wright}, {Griffith}, {Sigurdsson}, {Povich},
  \& {Mullan}}]{GHAT2}
{Wright}, J.~T., {Griffith}, R.~L., {Sigurdsson}, S., {Povich}, M.~S., \&
  {Mullan}, B. 2014, \apj, 792, 27, \dodoi{10.1088/0004-637X/792/1/27}

\bibitem[{{Youngblood} {et~al.}(2016){Youngblood}, {France}, {Loyd}, {Linsky},
  {Redfield}, {Schneider}, {Wood}, {Brown}, {Froning}, {Miguel}, {Rugheimer},
  \& {Walkowicz}}]{Youngblood2016}
{Youngblood}, A., {France}, K., {Loyd}, R.~O.~P., {et~al.} 2016, \apj, 824,
  101, \dodoi{10.3847/0004-637X/824/2/101}

\end{thebibliography}
\bibliographystyle{aasjournal}

%% This command is needed to show the entire author+affiliation list when
%% the collaboration and author truncation commands are used.  It has to
%% go at the end of the manuscript.
%\allauthors

%% Include this line if you are using the \added, \replaced, \deleted
%% commands to see a summary list of all changes at the end of the article.
%\listofchanges

\end{document}